# Imaging from the Visible to the Longwave Infrared wavelengths via an inverse-designed flat lens

*Monjurul Meem,* [1,*] *Sourangsu Banerji,* [1,*] *Apratim Majumder,* [1] *Juan C. Garcia,* [2] *Philip W. C. Hon,* [2]

*Berardi Sensale-Rodriguez* [1] *and Rajesh Menon* [1,3, a)]

[1]Department of Electrical and Computer Engineering, University of Utah, Salt Lake City, UT 84112, USA.

[2]Northrop Grumman Corporation, NG Next, Redondo Beach CA 90278, USA.

[3]Oblate Optics, Inc. San Diego CA 92130, USA.

a) rmenon@eng.utah.edu

[*] Equal contribution.

**ABSTRACT**

It is generally assumed that correcting chromatic aberrations in imaging requires optical elements. Here, we show that by allowing the phase in the image plane to be a free parameter, it is possible to correct chromatic variation of focal length over an extremely large bandwidth, from the visible (Vis) to the longwave infrared (LWIR) wavelengths using a single diffractive surface, *i.e.*, a flat lens. Specifically, we designed, fabricated and characterized a flat, multi-level diffractive lens (MDL) with thickness ≤ 10μm, diameter ~1mm, and focal length = 18mm, which was constant over the operating bandwidth of λ=0.45μm (blue) to 15μm (LWIR). We experimentally characterized the point-spread functions, aberrations and imaging performance of cameras comprised of this MDL and appropriate image sensors. We further show using simulations that such extreme achromatic MDLs can be achieved even at high numerical apertures (NA=0.81). By drastically increasing the operating bandwidth and eliminating several refractive lenses, our approach enables thinner, lighter and simpler imaging systems.

## Introduction

Imaging is a form of information transfer from the object to the image plane. This can be accomplished via a lens that performs a one-to-one mapping [1], via an unconventional lens (such as one with a structured point-spread function or PSF) that performs a one-to-many mapping, [2] or via no lens, where the light propagation performs a one-to-all mapping. In the first case, the image is formed directly. In the second case, the image is formed after computation, which can be especially useful, when encoding spectral [3,4], depth [5], polarization [6] or other information into the geometry of the PSF. Note that the modification of the PSF may be at the same scale as the diffraction limit [3-6] or it can even be much larger [7-10]. The image can be recovered in many cases in the no-optics scenario as well [11,12]. Intriguingly, machine learning may be employed to make inferences based on the acquired information (even without performing image reconstruction for human visualization), which has potential implications for privacy among others [13, 14].

The lens-based one-to-one mapping approach is preferred in many cases due to the high signal-to-noise ratio at each image point. When such a lens is illuminated by a plane wave, it forms a focused spot at a its focal plane. When imaging at optical frequencies, only the intensity is measured. As a result, the phase[1] of the field in the image or focal plane is a **free parameter**. Via back-propagation of the complex field to the lens plane, [15] we surmise that the lens-pupil function is, in fact, not unique. Here, we show that it is possible to search through all the possible (degenerate) lens-pupil functions to achieve achromatic focusing (and imaging) over a large operating bandwidth of λ=0.45μm to 15μm. We illustrate this via a single diffractive surface that is patterned with rings of width=8μm and heights varying from 0 to 10μm, fabricated within a polymer (positive-tone photoresist, see Fig. 1a). We refer to this diffractive surface as a multi-level diffractive lens (MDL). The focal length and diameter of the MDL are 18mm and 0.992mm, respectively. Figure 1a illustrates the phase shift imposed on an incident plane wave by the MDL at λ = 0.45 μm, which is expressed as $\Psi = (2\pi/\lambda)*h*(n-1)$, where h is the MDL-design height distribution and n is the refractive index at $\lambda$. The MDL pupil function is then expressed as $e^{-i\Psi}$. We note that the maximum ring height of 10μm corresponds to a phase shift much larger than $2\pi$ for many of the wavelengths in our design. The amplitude and phase of the field distribution in the focal plane is also plotted in Fig. 1a for λ=0.45μm, and in Fig. 1b for λ=5μm and 15μm. All image sensors measure the square of the amplitude of the field distribution, *i.e.*, the intensity distribution, and thereby, discard the phase. Here we show that a single diffractive surface can perform focusing and imaging that is substantially independent of wavelength over a range from Vis to LWIR, something that is considered impossible in conventional lenses [16]. Furthermore, contrary to popular belief, we demonstrate via simulations that this extreme bandwidth can be achieved even at high numerical apertures.

[1] In this paper we refer to phase of the scalar electromagnetic field averaged over many optical cycles as would be experienced by a sensor. In other words, the time-harmonic portion of the phase is not relevant.

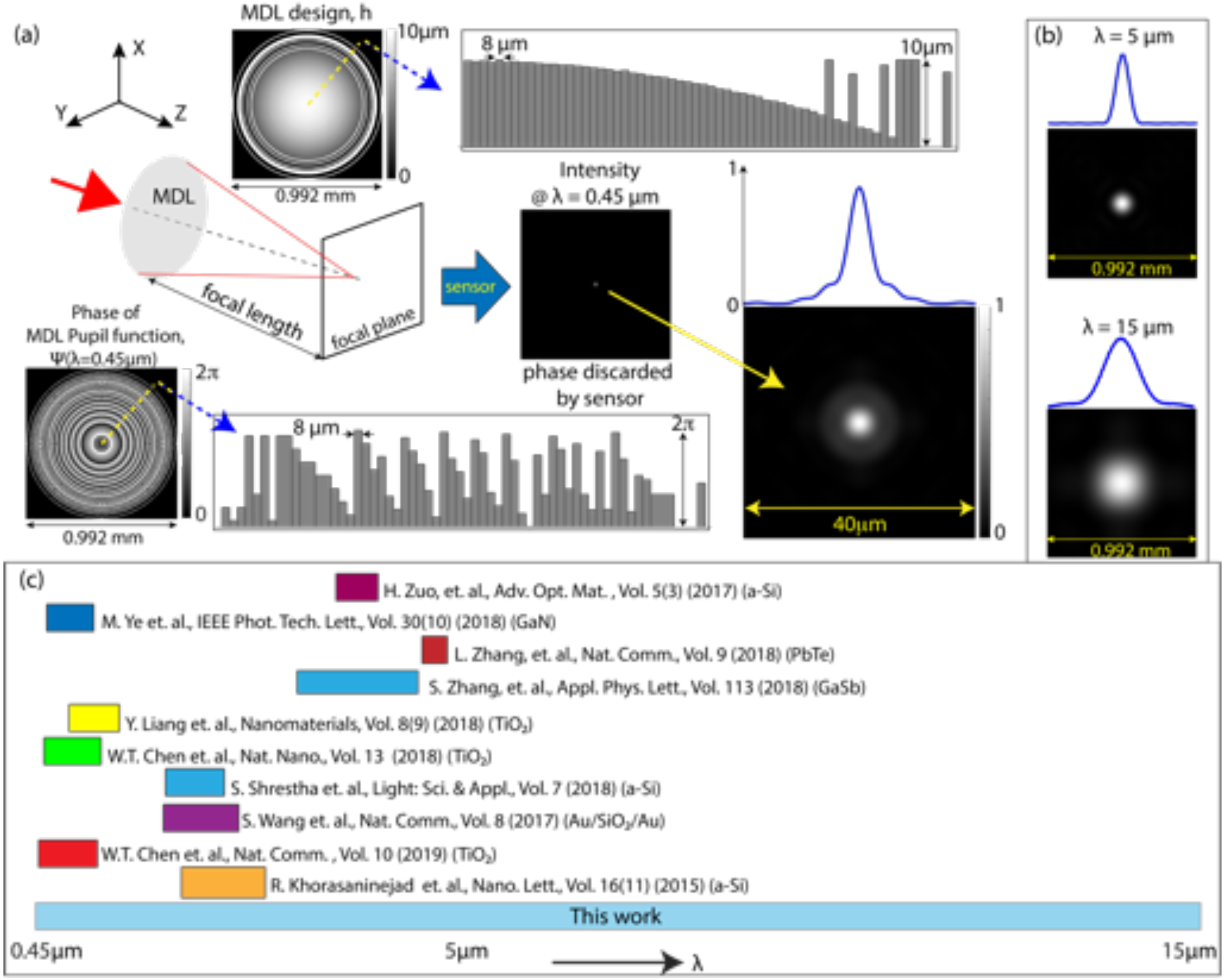

***Figure 1**: **Achromaticity with a single surface is achieved by allowing phase in the focal plane to be a free parameter.** (a) Schematic of a single surface multi-level diffractive lens (MDL) focusing collimated light to a focal spot. The MDL imparts its pupil function on the incident light. An image sensor discards the phase of the field in the focal plane, and only records the intensity. The height distribution (h) and the phase of the pupil function, $\Psi$ ($\lambda$=0.45μm) of a visible-LWIR MDL is shown. Note that the maximum h of 10μm is determined by fabrication and not the phase shift. The pupil function varies with $\lambda$. The normalized intensity distribution of the field in the focal plane at $\lambda$=0.45μm (top: cross-section through the center) is also shown. (b) The corresponding plots for $\lambda$=5μm and 15μm are shown. (c) Survey of operating bandwidths of reported metalenses to emphasize that our demonstration is almost 1 order of magnitude larger than any shown previously.*

It is important to note that in conventional refractive imaging systems, multiple lenses (sometimes made from different materials with differing dispersion properties) are used to correct for chromatic aberrations [16]. Not only are the individual refractive lenses thick and heavy, but multiple lenses require precise alignment during assembly. Metalenses have been proposed since the 1990s to mitigate these disadvantages [17, 18]. More recently, the parabolic phase profile with group-velocity compensation has been applied to correct for chromatic aberrations [19]. Via a careful literature study (see summary in Fig. 1c), we conclude that the largest bandwidth demonstrated via a metalens currently is 2μm, *i.e.*, from λ=3μm to 5μm [19]. However, the diameter of this metalens was only 30μm and it required ~200nm features in a high-refractive-index material (GaSb). Needless to say, no metalens to date has ever reported imaging over the visible to the LWIR bands. In fact, we recently showed that appropriately designed multi-level diffractive lenses (MDLs) are not only far easier to fabricate, but they outperform metalenses, and thereby concluded that metalenses do not offer any advantage for imaging [20]. We have already demonstrated separate MDLs operating in the visible (0.45μm to 0.75μm) [20-23], near infrared (0.845μm to 0.875μm) [24], short-wave infrared (0.875μm to 1.675μm) [25], long-wave infrared (8μm to 12μm) [26] and terahertz (1000μm to 3000μm) [27, 28] bands. In this paper, we show that a single MDL is actually able to image across the visible and the long-wave infrared bands, which we believe is the largest operating bandwidth for any flat lens demonstrated so far, and almost an order of magnitude larger than the largest bandwidth demonstrated before.

**Design**

The particular details of our design methodology is similar to what has been reported previously [20-28]. To summarize, we maximize the wavelength-averaged focusing efficiency of the MDL by selecting the distribution of heights of the rings that form the MDL. This selection is based upon a gradient-assisted direct-binary-search technique. We included a constraint of at most 100 height levels with a maximum individual height level of 10μm and minimum feature width of 8μm, where were all dictated by our fabrication process. The material dispersion of a positive-tone photoresist, AZ9260 (Microchem) was assumed [26]. The height distribution of the designed MDL is shown in Fig. 1a (bottom-left inset), while the phase transmittance function (lens pupil function) at $\lambda$=0.45μm is shown in the center inset. As mentioned earlier, when this field is propagated to the focal plane, 18mm away from the MDL, the resulting amplitude and phase distributions as well as the intensity distributions are also shown in Fig. 1a. The corresponding plots for $\lambda$=5μm and $\lambda$=15μm are shown in Fig. 1b. Even though the phase distribution in the focal plane differs between the wavelengths, the intensity distributions are almost identical and simply scale with wavelength as expected, resulting in a single-surface lens that is achromatic from 0.45μm to 15μm. The simulated PSFs for all the wavelengths are depicted in Fig. S1 [29].

**Experiments**

The MDL was fabricated using grayscale lithography as reported previously [19-26]. We first utilized CVD-diamond as the support substrate as it is relatively transparent from the visible to the LWIR (see transmission spectrum in Fig. S2) [29]. However, the polycrystalline grains of the CVD diamond scatter light too much reducing the efficiency of the lenses (Figs. S3-S5). Therefore, we fabricated a pair of lenses of the same design: one on a glass wafer for $\lambda$=0.45μm to ~2μm and another on a silicon wafer for $\lambda$ ~2μm to 15μm. We emphasize that the patterned polymer was identical for both lenses. Optical micrographs of these MDLs are shown in Fig. 2a. The MDL on a glass wafer was placed in front of a silicon monochrome image sensor (DMM 27UP031-ML, Imaging Source) for visible and near-IR imaging. The average transmission efficiency in the visible and NIR wavelengths of the MDL with glass substrate was ~85%. The point-spread function (PSF) of the MDL was measured by illuminating it with a collimated beam from a tunable supercontinuum source (NKT Photonics SuperK Extreme with SuperK VARIA filter for visible wavelengths, 0.35μm-0.85μm and SuperK SELECT filter for near infrared wavelengths, 0.8μm-0.9μm) (Fig. S6) [20-25]. The wavelength of the source was tuned from 0.45μm to 0.85μm in steps of 50nm and bandwidth of 10nm. The focused spot at each wavelength was relayed with magnification (22.22X) onto the image sensor. The simulated and captured raw images from λ=0.45μm to 0.9μm are shown in Figs. 2b and 2c. The infrared experiments were performed by placing the silicon-substrate MDL in front of a BST microbolometer focal-plane array (PV320L, Electrophysics Scientific Imaging). The MDL was illuminated by collimated beam from a supercontinuum fiber source (SuperK Extreme/Fianium, NKT Photonics) for λ=1.05μm and 2μm (continuous wave), or by a tunable quantum cascade laser (MIRcat-QT$^{TM}$, DRS Daylight Solutions) for λ=4μm to 11μm (each with a pulse width of 100ns). The simulated and measured point-spread functions for representative wavelengths in the SWIR, MWIR and LWIR bands are shown in Figs. 2d-2f. We emphasize that the focal length is 18mm for all wavelengths from 0.45μm to 15μm.

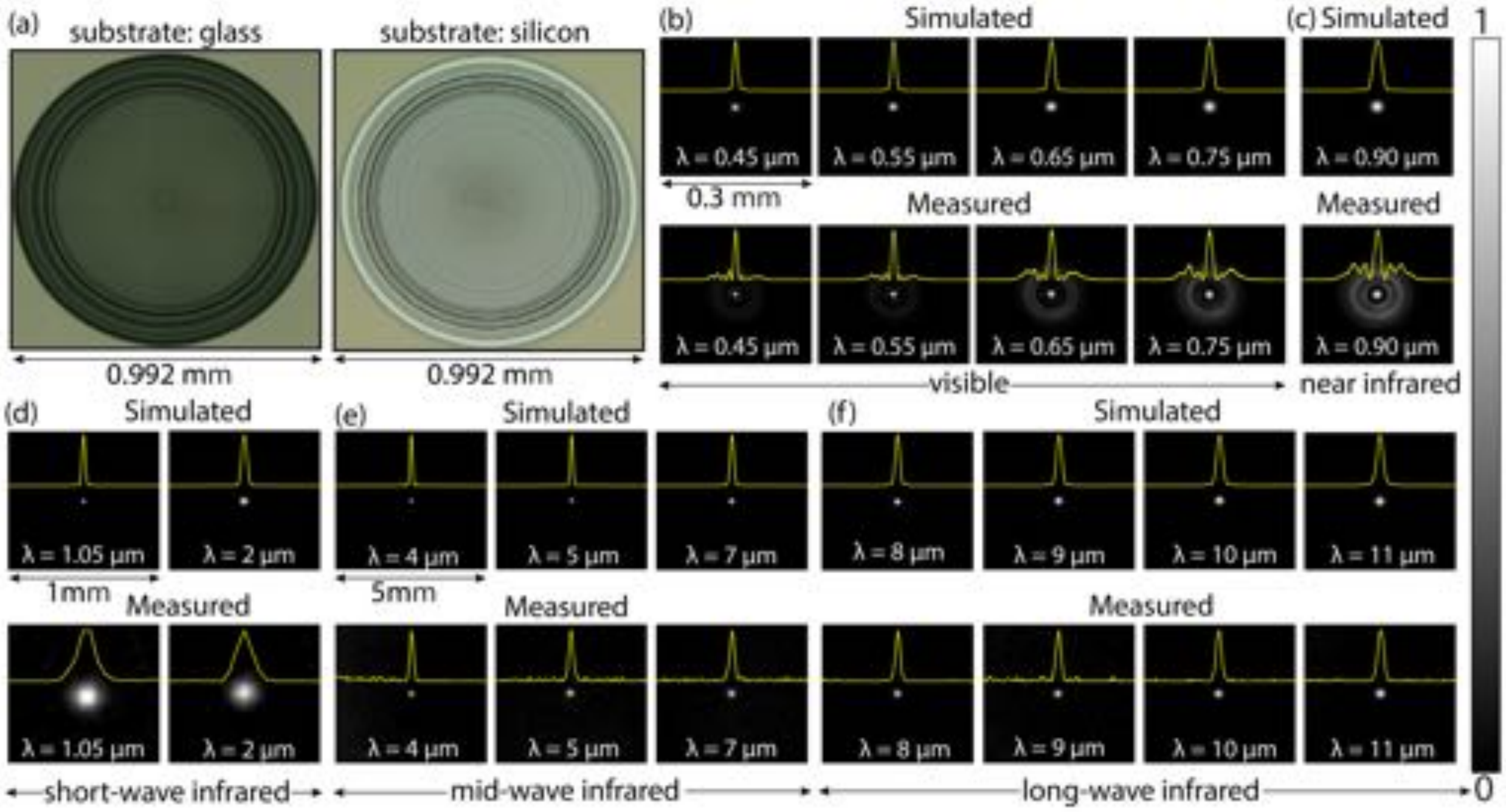

***Figure 2: Achromatic focusing by the MDL.*** *(a) Optical micrographs of the fabricated MDLs. The device is the same, but the substrate is glass for visible and NIR, while the substrate is silicon for the other wavelengths. Simulated and Measured point-spread functions at the same focal plane, 18mm away from the MDL in (b) the visible spectrum and (c-f) the infra-red spectrum. A silicon CMOS image sensor and glass-substrate MDL were used for the (b) visible and (c) NIR images. A BST microbolometer focal-plane array and Si-substrate MDL were used for all images at λ>1μm (d-f). The size of each panel is 0.3mm,*

*0.3mm, 1mm, 5mm and 5mm in (b), (c), (d), (e) and (f), respectively. The sizes for the simulation and experiment results are the same.*

We first note that there is wide discrepancy in the reporting of focusing efficiencies in flat lenses. Here, we define the he focusing efficiency of the MDL as the power within a spot of diameter equal to 3 times the FWHM divided by the total power incident on the lens [20]. The simulated wavelength averaged focusing efficiency for the designed MDL is 91% in comparison to a measured value of ~32% in the 0.45μm to 15μm band (Fig. S7) [29]. In order to explain this discrepancy, we performed careful simulations of the sensitivity of the focusing efficiency to errors in the ring heights (Fig. S8) [29]. The analysis suggests that standard deviation in ring heights of 0.4μm can explain the drop in efficiency. Such standard deviations are expected in our existing grayscale lithography process (see Fig. S9 for error measurements) [29]. Fabrication errors in the pixel width (Fig. S10) as well as combination of width and height errors (Figs. S11, S12) are also possible explanations. Simulations suggest that errors in the refractive index of the polymer could also explain some discrepancies (Fig. S13). In the future, it is possible to incorporate tolerance to fabrication errors as one of the metrics during the optimization-based design step, analogous to what was done previously for binary multi-wavelength diffractive lenses [30]. The diffraction-limited, measured and simulated full-width at half-maxima (FWHM) as function of wavelength are plotted in Fig. 3a. It is noted that besides in the SWIR and MWIR bands, the measured spot is close to the diffraction limit.

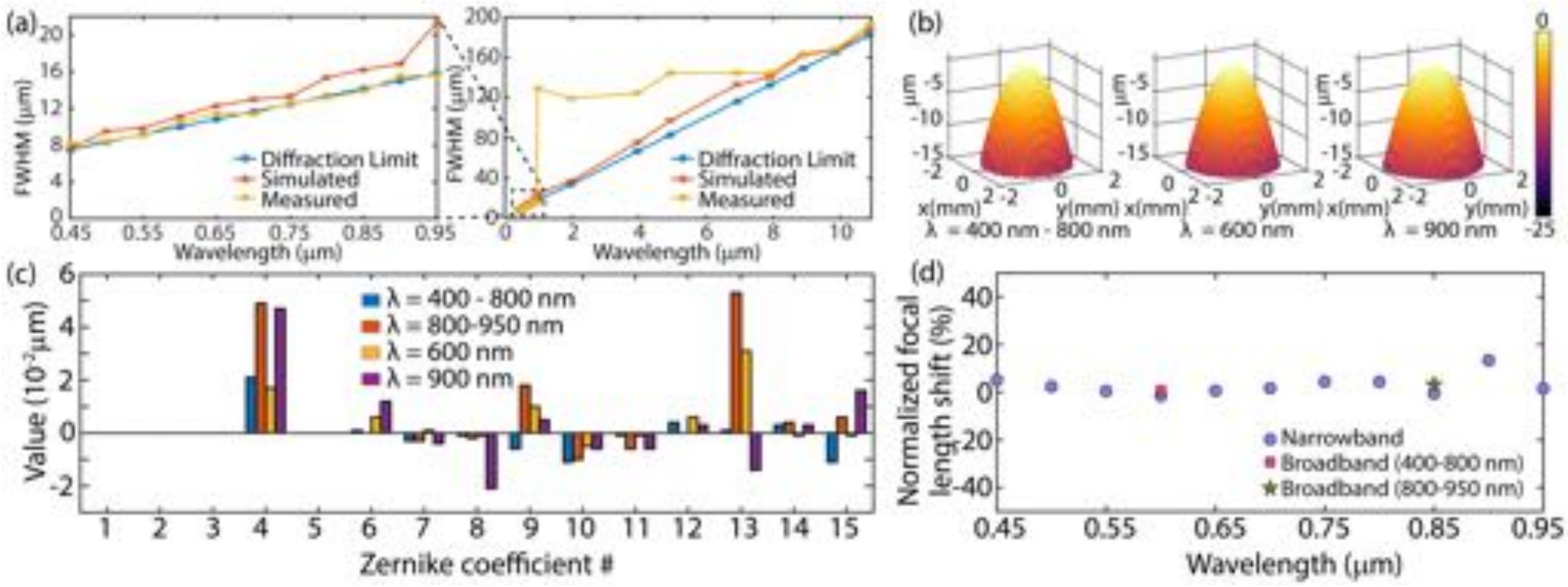

***Figure 3**: **Analysis of the MDL.** (a) Measured and simulated FWHM of the PSF as a function of wavelength. Except in the SWIR and MWIR bands, the measured FWHM is close to the simulated and diffraction-limited values. (b) Measured reconstructed wavefronts under broadband (0.4μm to 0.8μm), and*

*narrowband (0.6µm, 0.9µm) illuminations. (c) Measured aberrations coefficients for both broadband and narrowband illuminations. (d) Measured shift in focal length as a function of wavelength (Vis-NIR). The average shift is 3%.*

The wavefront aberrations of the glass-substrate MDL were measured using a Shack-Hartmann wavefront sensor (Thorlabs, WFS 150-7AR) [29]. The wavefront aberrations were measured under broadband (0.4µm-0.8µm for the visible spectrum and at 0.8µm-0.95µm for the NIR spectrum) and under narrowband illuminations at 0.45µm, 0.5µm, 0.55µm, 0.6µm, 0.65µm, 0.7µm, 0.75µm and 0.8µm with 50nm bandwidth and 0.85µm, 0.9µm and 0.95µm with 15nm bandwidth. Most importantly, the reconstructed wavefronts are quite similar for both the broadband and narrowband illuminations (Fig. 3b), confirming excellent achromaticity. The corresponding Zernike polynomial coefficients for measurements under 0.4µm-0.8µm, 0.8µm-0.95µm, 0.6µm and 0.9µm are shown in Fig. 3c. The measurements confirm that the MDL indeed has low values for all aberrations (Table S1) [29]. Following a well-known procedure that calculates the focal length using the radius of curvature (ROC) measured from the WFS and the recorded distance between the test lens and the WFS, we calculated the change in focal length from the nominal value of 18mm as a function of wavelength (Fig. 3d) (details in section S2), confirming good achromatic behavior [29].

Next, we assembled a camera by placing the MDL in front of the appropriate image sensor. We characterized the imaging behavior of the MDL by capturing still and video images of various objects. The results are summarized in Fig. 4. See Supplementary Videos 1-4 for videos using the MDL with glass substrate under sunlight with NIR-cut filter, sunlight with vis-cut filter, white-LED light and with 0.85µm-LED flashlight, respectively, and Supplementary Video 5 using the MDL with Si substrate of a heated resistor coil. Note that corresponding visible, NIR and LWIR videos using the MDL with diamond substrate are in Supplementary Videos 6-8 (Table S2) [29]. The field of view of the MDL is estimated as ~15$^0$. For majority of images, the distance between the MDL and the sensor was ~19mm for all wavelengths, and the distance between the MDL and the object was 450mm for the visible and NIR bands, and 425mm for the IR bands. In each case, the exposure time was adjusted to ensure that the frames were not saturated. In addition, a dark frame was recorded and subtracted from the images. For the images in the SWIR and longer wavelengths, a hot-plate at temperature of approximately $200^0$C was placed behind the objects in order to image their silhouettes. The only exceptions to this were the images of the soldering iron (3rd column in Fig. 4 at ~$180^0$C) and the heated resistor coil (only LWIR, 5th column in Fig. 4 at ~$150^0$C). The resolution-chart images in visible and NIR (Fig. 4) shows that the resolved spatial frequency is ~28.5 line-pairs/mm, which corresponds to a spatial period of ~17.5μm or about ~8 times the sensor pixel size (2.2μm). This resolution corresponds approximately to the average FWHM over the visible-NIR bands (Fig. 3a). We also note that the visible image of the Macbeth color chart shows excellent color reproduction.

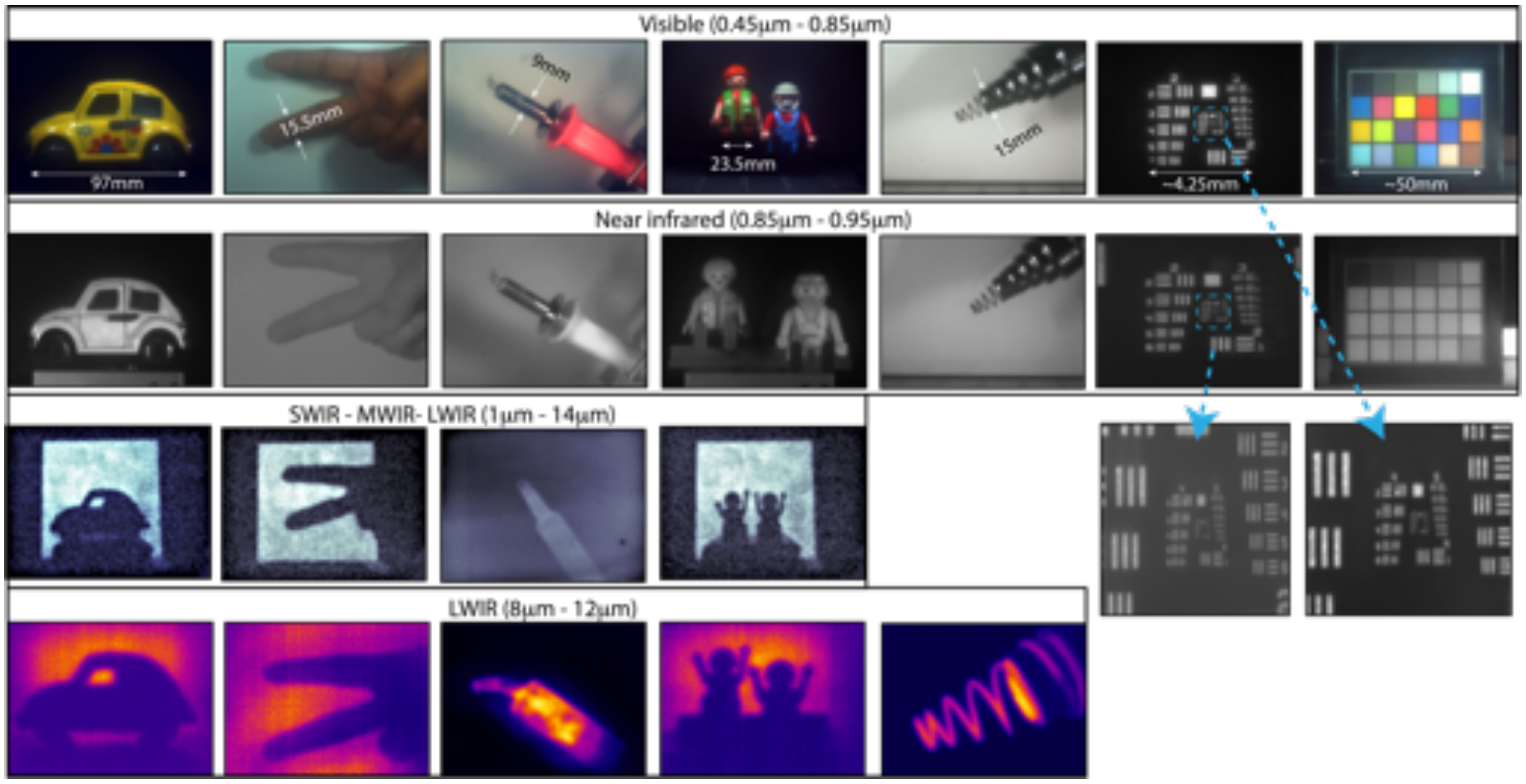

***Figure 4: Imaging from the visible to the LWIR using a single MDL.*** *The visible and NIR images were taken using the silicon CMOS image sensor. The LWIR images were taken using a FLIR Tau2 camera sensor. The remaining images used the BST microbolometer focal plane array. The MDL with glass substrate was used for visible and NIR wavelengths, while the MDL with Si substrate was used for all other wavelengths. Note that all images are raw and unprocessed. Also see corresponding videos in Supplementary Videos 1-8. Note that for the SWIR-MWIR-LWIR images, the lower wavelength cut-off arises from transmission of the silicon substrate, and no filters were used.*

**Extreme Achromaticity at high Numerical Aperture**

There is a misconception that MDLs cannot operate at high numerical apertures [31]. This is completely incorrect, as we have pointed out previously [20]. Here, we further show that extreme operating bandwidth of 0.45μm to 15μm can be achieved even at a numerical aperture (NA) as large as 0.81. Specifically, we designed an MDL with diameter=25μm, focal length=9μm and NA=0.81 as illustrated in Fig. 5a. The maximum height of the MDL was 5μm with up to 128 height levels and ring width was 0.225μm. The PSFs of the MDL were computed using finite-difference time-domain (FDTD, see methods). The simulated focusing efficiency averaged over all the wavelength is 78% (as shown in Fig. 5b). Similarly, the FWHM of the PSFs are close to the diffraction limit (Fig. 5c). The simulated intensity distributions at exemplary wavelengths spanning the operating bandwidth in the XZ planes are shown in Figs. 5(d-t) and confirm that the focal length is independent of wavelength. Note that all simulations are performed assuming unpolarized light. This study confirms that the concept of phase in the focal plane as a free parameter can be readily applied to enable extreme achromaticity in high-NA MDLs.

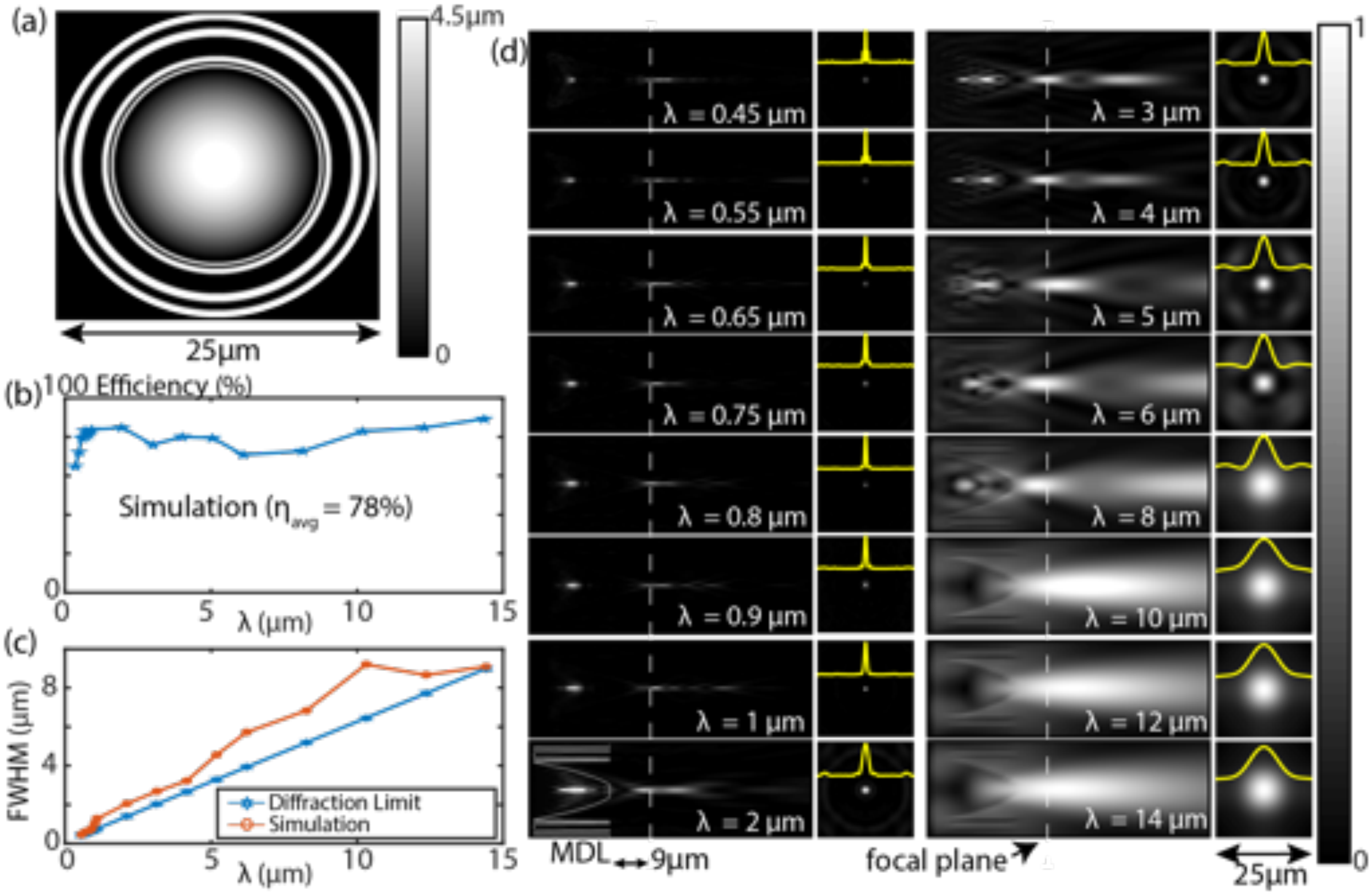

***Figure 5: Extreme achromaticity in high-NA MDL.*** *(a) Design of an MDL with f=9μm, diameter=25μm and λ=0.45μm to 15μm. Simulated (b) efficiency and (c) FWHM as a function of the illumination wavelength. (d) Simulated focal intensity distributions in the XZ plane (right insets show top-view or XY plane) for exemplary wavelengths spanning the operating bandwidth. Note that the outline of the MDL is indicated by white lines in the bottom-left panel. The dashed white lines indicate location of*

*the focal plane. All XZ planes simulations have the same size, while all XY planes simulations have the same scale of 25μm.*

## Conclusions

The most important conclusion of this work is that an ideal lens does not possess a *unique* pupil function. In other words, the ideal lens need not have a parabolic phase profile as is commonly understood. By removing this restriction, we enable numerous solutions to the ideal lens problem. Then, the final choice can be made based upon other requirements such as achromaticity (as we described here), extreme depth of focus [32], manufacturability, or minimization of aberrations, of weight, of thickness, of cost, etc. Our approach can be readily generalized to metasurfaces (by employing full-wave diffraction models), which could be advantageous to manipulate the polarization states of light [33] and could even be implemented in integrated-photonics platforms [34-36]. The next obvious question is what limits the bandwidth of such an MDL. Initial simulations indicate that the bandwidth can be far larger than what we demonstrated here [37]. The space-bandwidth product (SBP) of the MDL, which is defined as the number of independent degrees of design freedom (lens-radius/ring-width*number of height levels) is expected to be the key limiting factor [38]. As per the channel-capacity theorem, we would expect that larger the SBP, the larger the operating bandwidth for a given field of view. Another important application for extreme bandwidth flat lenses would be in imaging and focusing of ultra-short pulses (which have large spectral bandwidths).

## Methods

**Modeling and Design:** The field in the focal plane, $U(x',y',\lambda, d)$ of the MDL is modelled with the Fresnel-Kirchhoff diffraction integral as:

$$U(x',y',\lambda,d) = \frac{e^{ikd}}{i\lambda d}\iint T(x,y,\lambda).\, e^{i\frac{k}{2d}[(x-x')^2+(y-y')^2]}dxdy, \quad (1)$$

where $d$ is the focal length, (x,y) are coordinates in the MDL plane, (x',y') are coordinates in the focal plane, $and\ T(x,y,\lambda)$ is the MDL pupil function. The intensity in the focal plane is given by $I(x',y',\lambda,d) = |U(x',y',\lambda,d)|^2$. We can express equation (1) as $U(x',y',\lambda,d) = P\{T(x,y,\lambda)\}$, where P{} is an operator that transforms the pupil function into the field in the focal plane. We note that $P$ is analytic and the integral is bounded by the finite spatial extent of the pupil function. It is well known that $P$ is invertible [14]. In other words, the focal-plane field may be back-propagated to the MDL plane. This is evident, since Maxwell's equations are time reversible. Therefore, we can express the pupil function as

$$T(x,y,\lambda) = P^{-1}\{U(x',y',\lambda,d)\} = P^{-1}\{A(x',y',\lambda,d)e^{iB(x',y',\lambda,d)}\}, \quad (2)$$

where $A$ and $B$ are real-valued functions representing the amplitude and phase of the complex scalar field in the focal plane, respectively. In designing a lens, the only restriction we need is $A$ = sqrt ($I_{des}$), where $I_{des}$ is the desired intensity distribution of the focal spot. From (2), it's clear that there will be one function representing $T$ for each choice of $B$ such that the field U is a valid solution to Maxwell's equations. **Therefore, *T*, the pupil function of a lens, is not unique.**

A somewhat trivial example of this non-uniqueness is the following. A parabolic-phase pupil function will convert a plane wave to a converging spherical wave. However, one can add any arbitrary function riding on this parabola whose spatial frequencies are larger than those that will propagate in free space, without having any effect on the focal spot (assuming that the focal length >> λ). In other words, one can modify the pupil function in infinitely different ways without modifying the focal spot.

A more general example can be gleaned from holography. A hologram is designed to project a certain intensity pattern, and allow phase in the image plane to be arbitrary. There are multiple degenerate hologram phase functions that give the same image intensity distribution.

For design purposes, we utilized a modified version of direct-binary search to optimize the height profile of rotationally symmetric MDLs consisting of individual constituent rings of width equal to a pre-defined value with the explicit goal of maximizing wavelength-averaged focusing efficiency. This is equivalent to searching for an optimal function $B$ that satisfies the constraints placed on the pupil function, $T$ (phase only and space-bandwidth product limited by fabrication), while maintaining $A = \text{sqrt}(I_{des})$.

**FDTD Simulation:** The point-spread function of the high-NA MDL was simulated using the full 3D finite-difference time-domain (FDTD) method with the incident electric field polarized in the plane of the MDL. We averaged the fields over the two orthogonal polarization directions of the electric field in order to simulate the point-spread function (PSF) under unpolarized light. All analysis in the main text utilized this PSF assuming unpolarized input. In the FDTD simulations of the MDL, the entire region from the back surface of the lens up to 1.5 times the focal length was considered. Perfectly matched layer (PML) boundary conditions were placed around the computational region. The mesh accuracy in the FDTD software was set to $\sim\lambda/20$.

**Fabrication and Characterization:** The MDL height distribution was patterned in a photoresist (AZ9260) film atop a glass wafer using grayscale laser patterning using a Heidelberg Instruments MicroPG101 tool. The exposure dose was varied as a function of position in order to achieve the multiple height levels dictated by the design. After fabrication, the devices were characterized on an optical bench by illuminating them with broadband collimated light, whose spectral bandwidth could be controlled by a tunable filter.

**Acknowledgements**

We thank Brian Baker, Steve Pritchett and Christian Bach for fabrication advice, and Tom Tiwald (Woollam) for measuring dispersion of materials. RM acknowledges useful discussion with Henry Smith and Fernando Vasquez-Guevara. We also acknowledge a grant of credit on Amazon AWS (051241749381) for computation. The support and resources from the Center for High Performance Computing at the University of Utah as well as funding from the Office of Naval Research grant N66001-10-1-4065, NSF CAREER award: ECCS #1351389, NSF ECCS #1828480 and NSF ECCS #1936729 are gratefully acknowledged.

**Competing Interests Statement**

RM is co-founder of Oblate Optics, Inc., which is commercializing technology discussed in this manuscript. The University of Utah has filed for patent protection for technology discussed in this manuscript.

**Author Contributions**

RM, BSR, MM and SB conceived and designed the experiments. SB and RM modeled and optimized the devices. MM fabricated the devices. MM, SB and AM performed the Vis and NIR experiments. SB performed the MTF analysis. PWCH, JCG, MM and RM performed PSF measurements and imaging experiments in the IR. SB and AM performed the aberrations analysis. All authors performed the data analysis and edited the manuscript.

**Materials and Correspondence**

Correspondence and materials requests should be addressed to RM at rmenon@eng.utah.edu.

# Supplementary Material:

# Imaging from the Visible to the Longwave Infrared wavelengths via an inverse-designed flat lens

*Monjurul Meem,* [1,*] *Sourangsu Banerji,* [1,*] *Apratim Majumder,* [1] *Juan C. Garcia,*[2] *Philip Hon,*[2] *Berardi Sensale-Rodriguez*[1] *and Rajesh Menon*[1,3, a)]

[1]Department of Electrical and Computer Engineering, University of Utah, Salt Lake City, UT 84112, USA.

[2]Northrop Grumman Corporation, NG Next, Redondo Beach CA 90278, USA.

[3]Oblate Optics, Inc. San Diego CA 92130, USA.

a) rmenon@eng.utah.edu

[*] Equal contribution.

## 1. Simulations and Experiments of PSFs

The simulated PSFs for the MDL design are as follows:

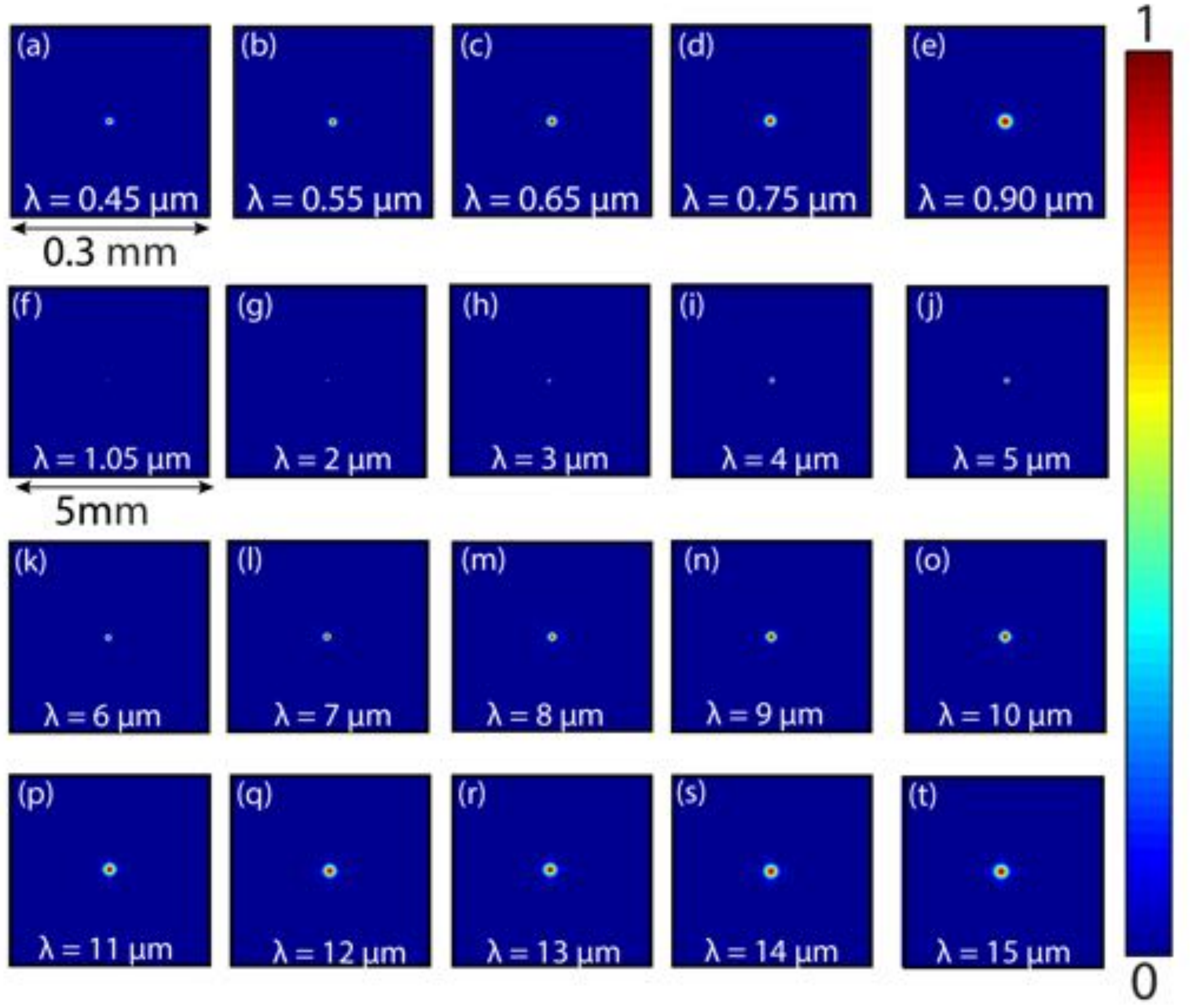

***Fig. S1:*** *Simulated PSFs for (a) 0.45 µm (b) 0.55 µm (c) 0.65 µm (d) 0.75 µm (e) 0.90 µm (f) 1.05 µm (g) 2 µm (h) 3 µm (i) 4 µm (j) 5 µm (k) 6 µm (l) 7 µm (m) 8 µm (n) 9 µm (o) 10 µm (p) 11 µm (q) 12 µm (r) 13 µm (s) 14 µm and (t) 15 µm*

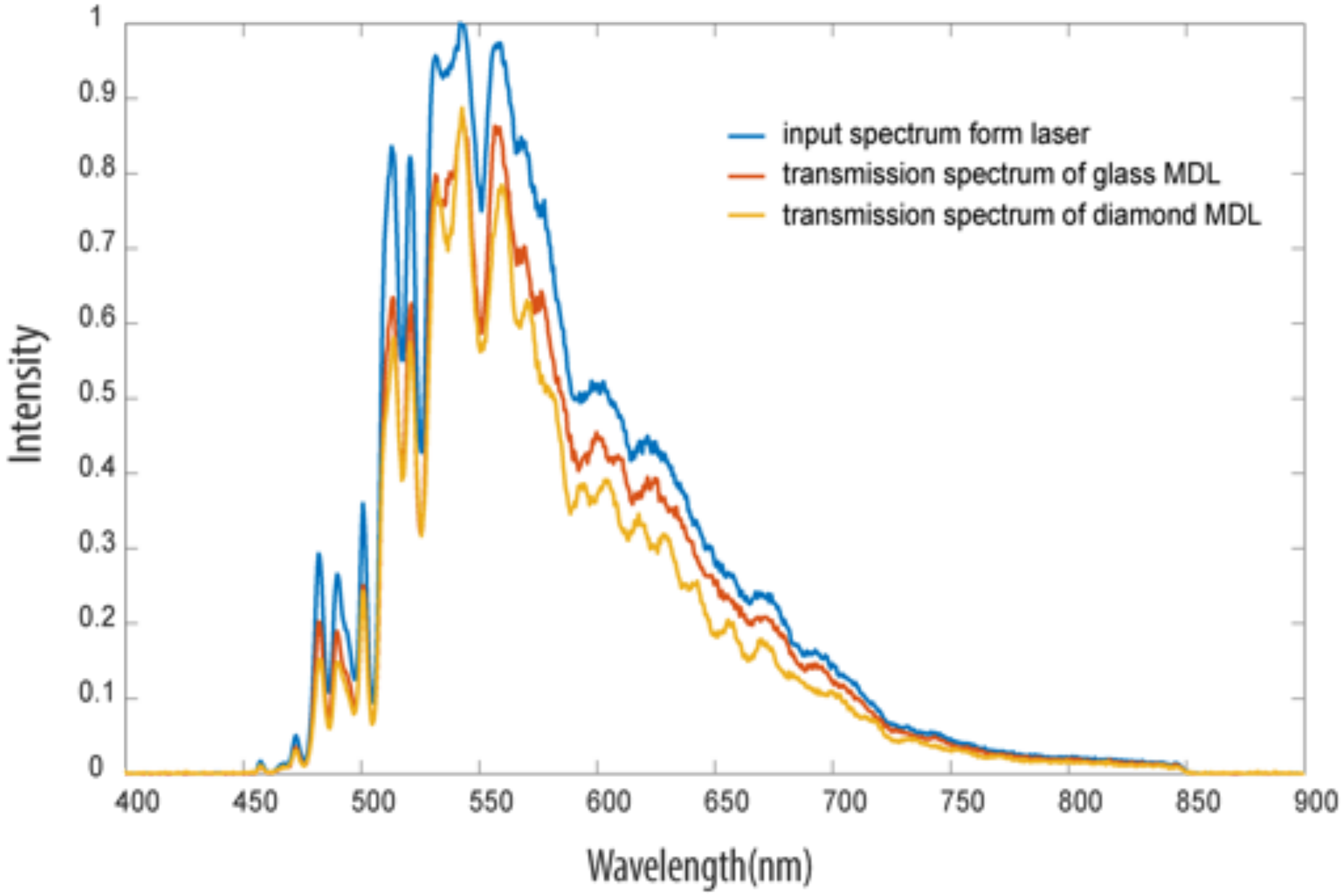

***Fig. S2:*** *Transmission spectra of all glass and diamond MDLs in visible (450nm – 850nm).*

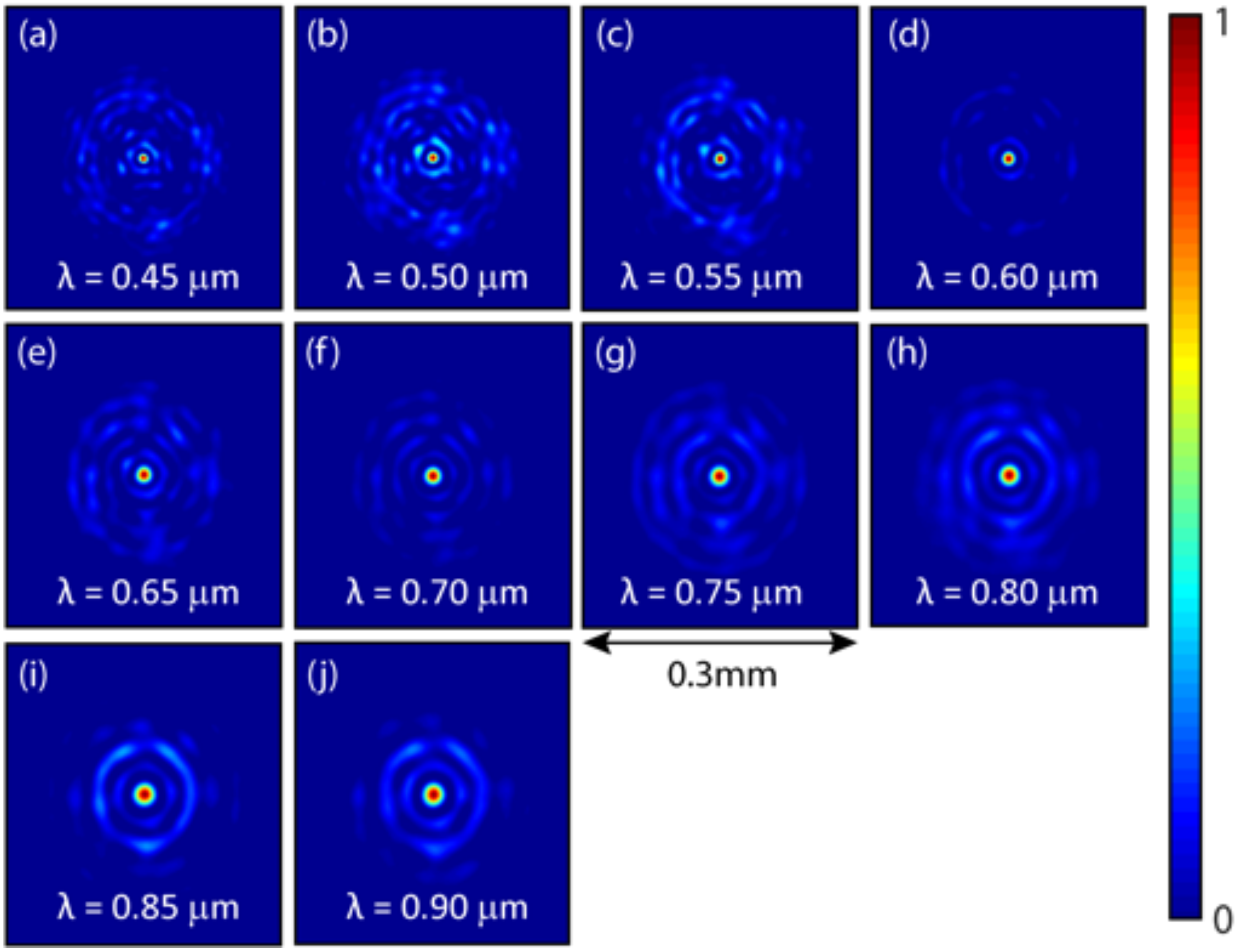

***Fig. S3:*** *Measured PSFs of the MDL on diamond substrate for (a) 0.45 µm (b) 0.50 µm (c) 0.55 µm (d) 0.60 µm (e) 0.65 µm (f) 0.70 µm (g) 0.75 µm (h) 0.80 µm (i) 0.85 µm (j) 0.90 µm.*

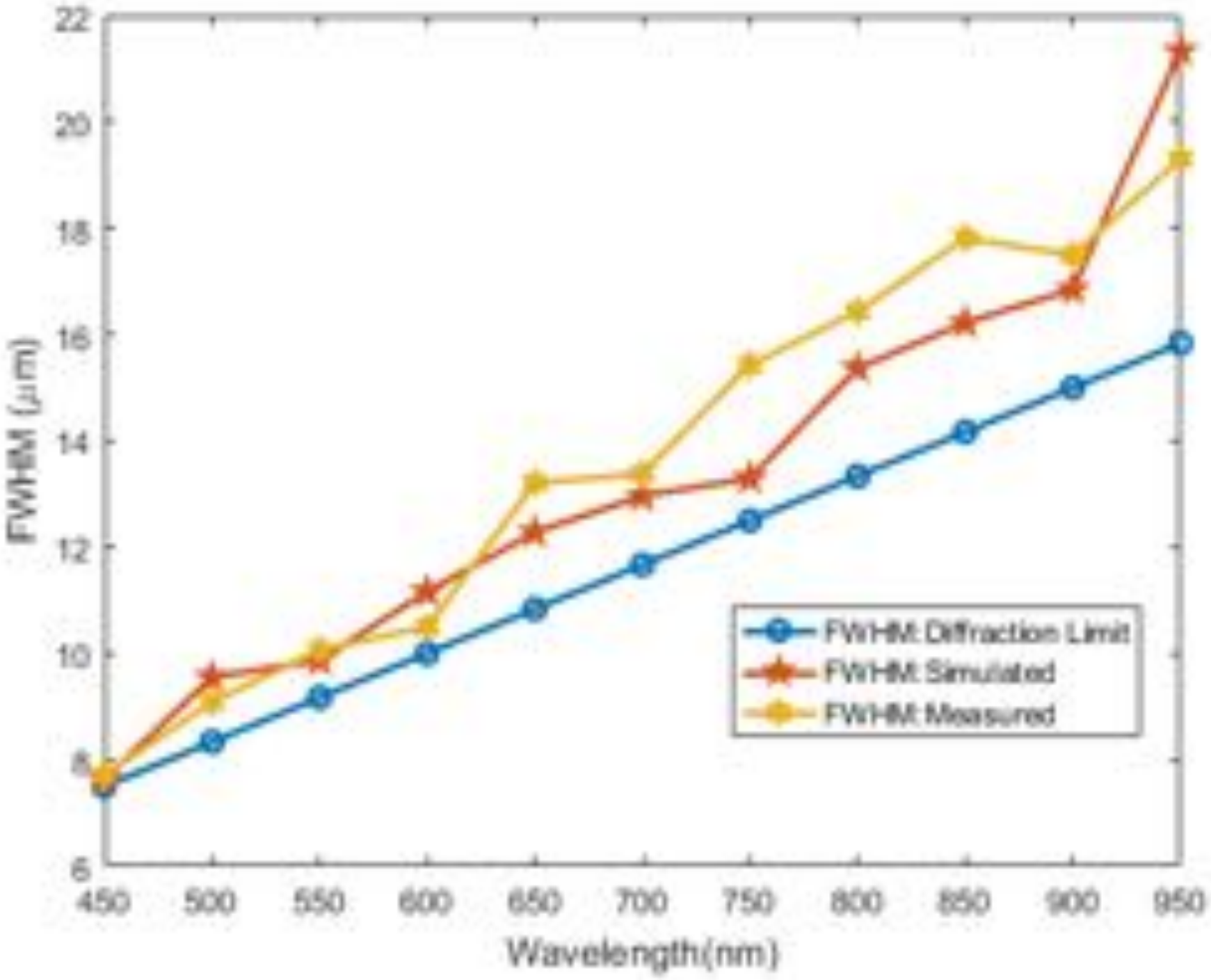

***Fig. S4:*** *Full Width Half Maximum (FWHM) of the MDL on diamond substrate*

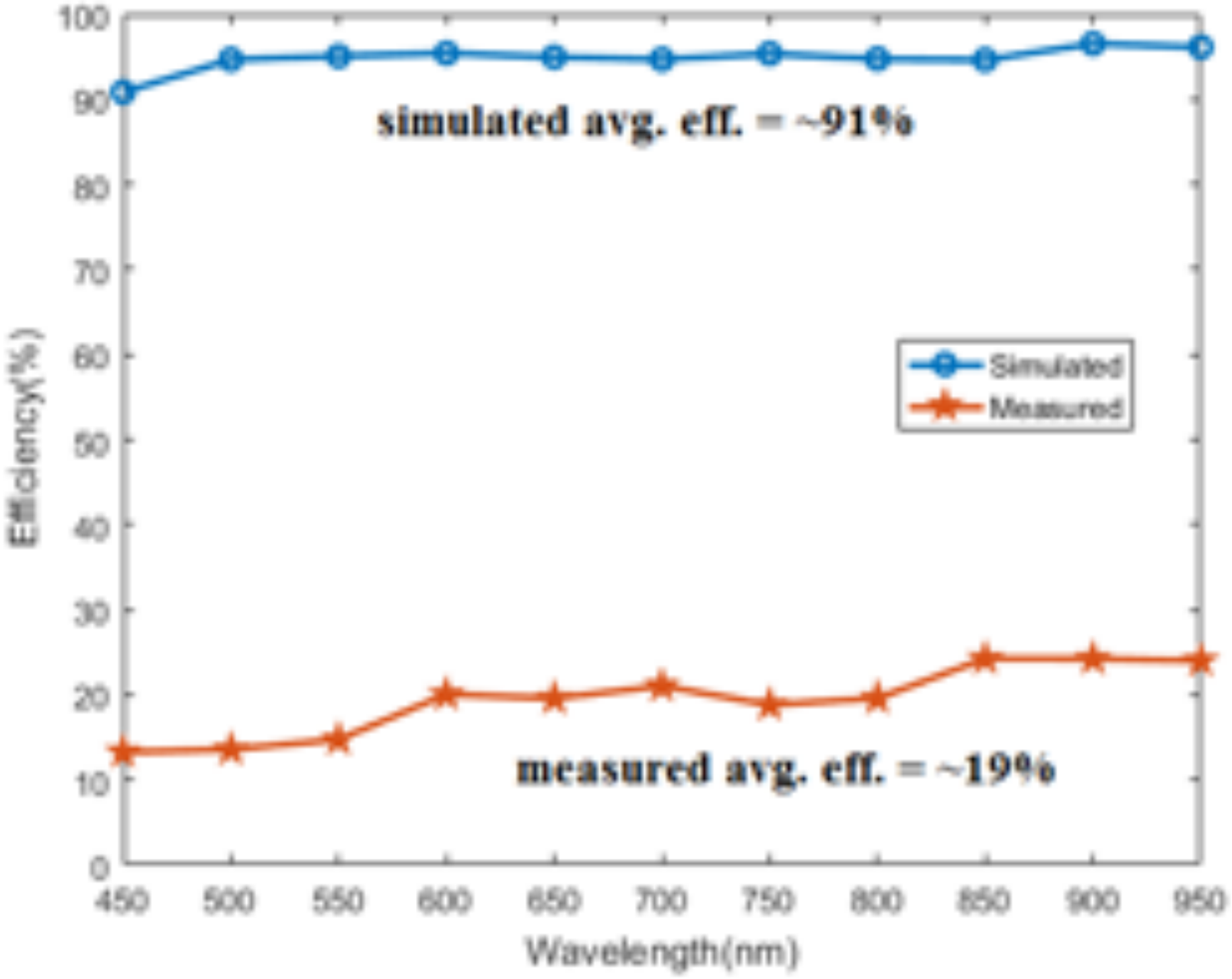

***Fig. S5:*** *Efficiency of the MDL on diamond substrate*

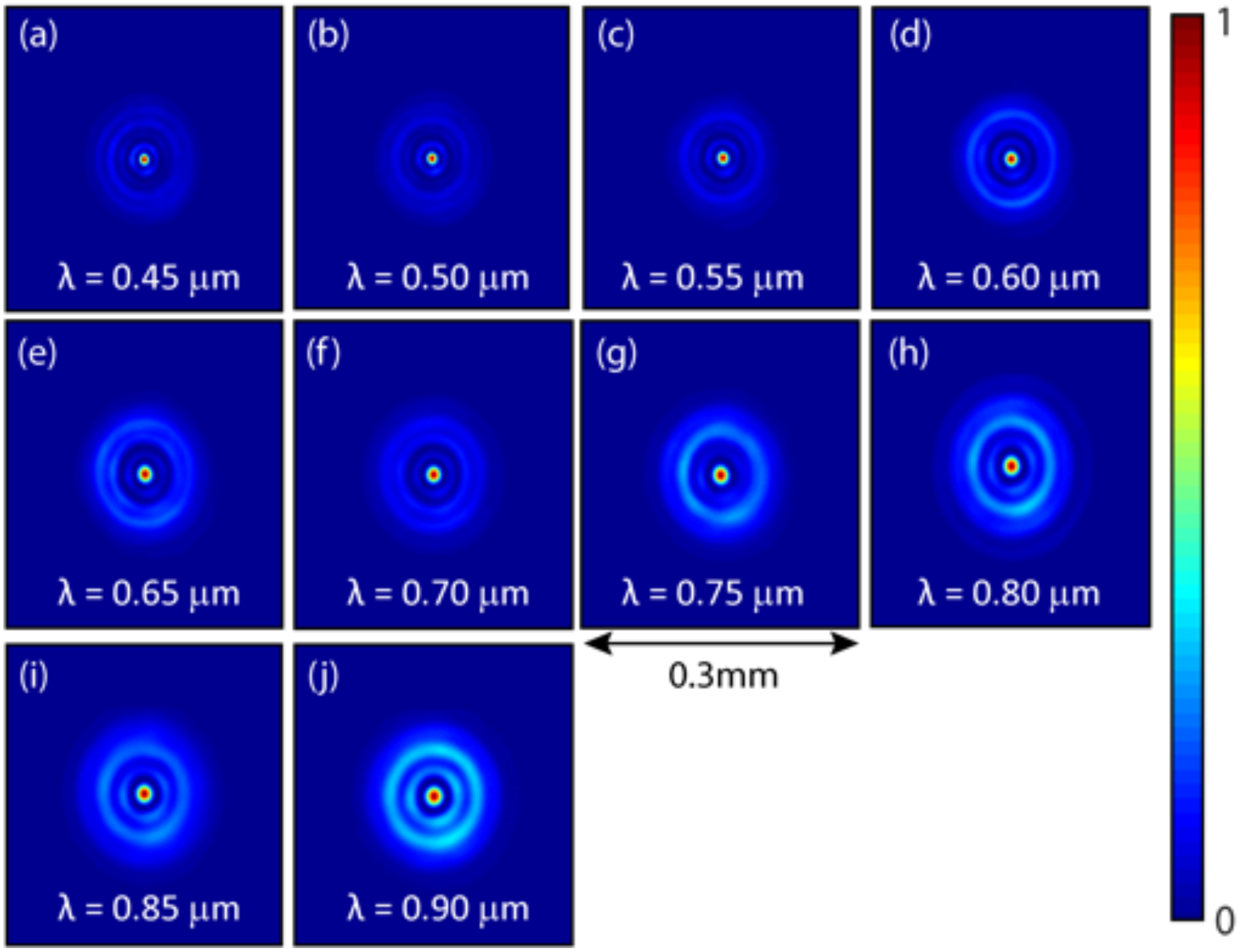

***Fig. S6:*** *Measured PSFs of the MDL on glass substrate for (a) 0.45 μm (b) 0.50 μm (c) 0.55 μm (d) 0.60 μm (e) 0.65 μm (f) 0.70 μm (g) 0.75 μm (h) 0.80 μm (i) 0.85 μm (j) 0.90 μm.*

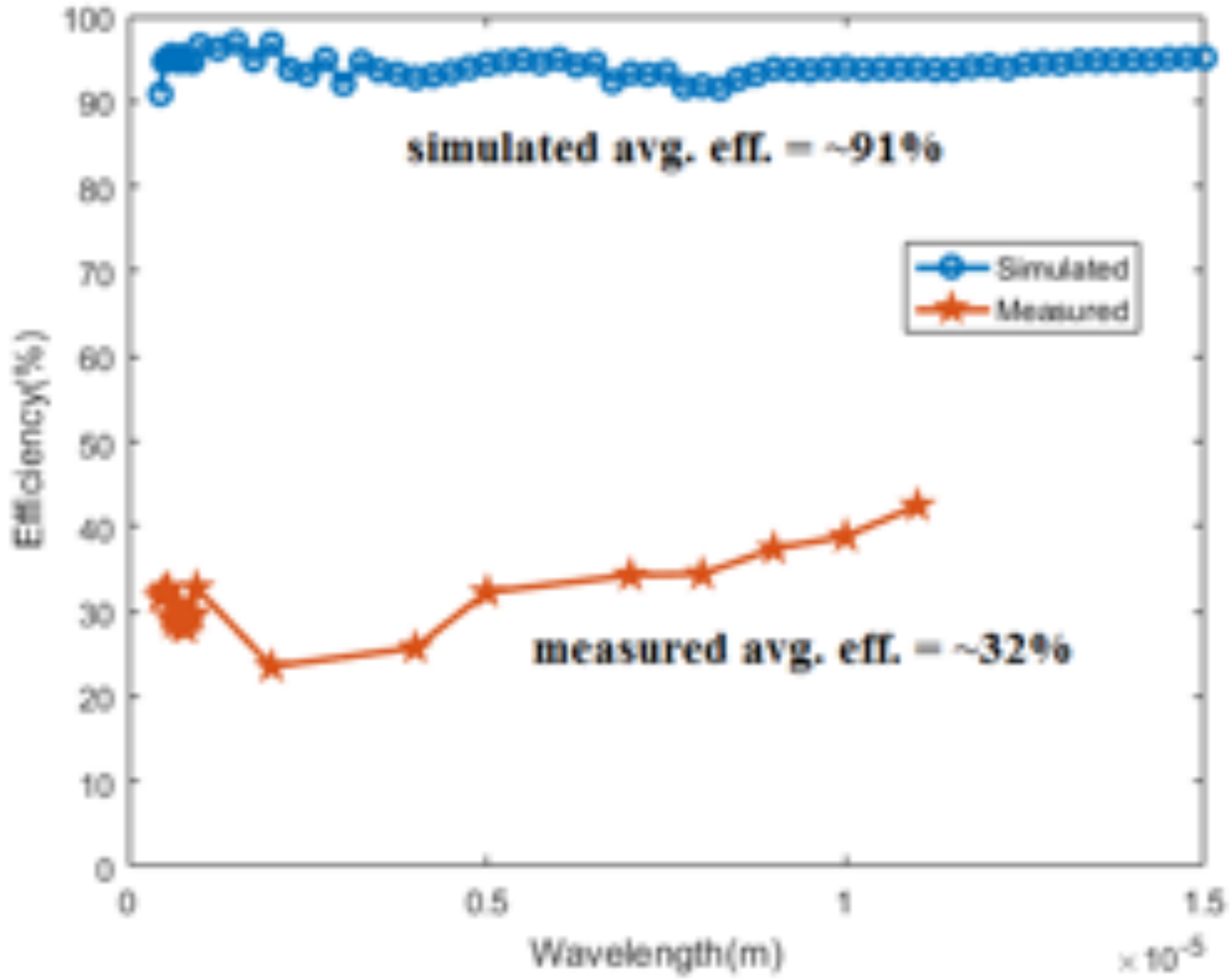

***Fig. S7:*** *Efficiency spectrum of the MDL. Glass substrate was used for visible and NIR, while Si substrate was used for longer wavelengths.*

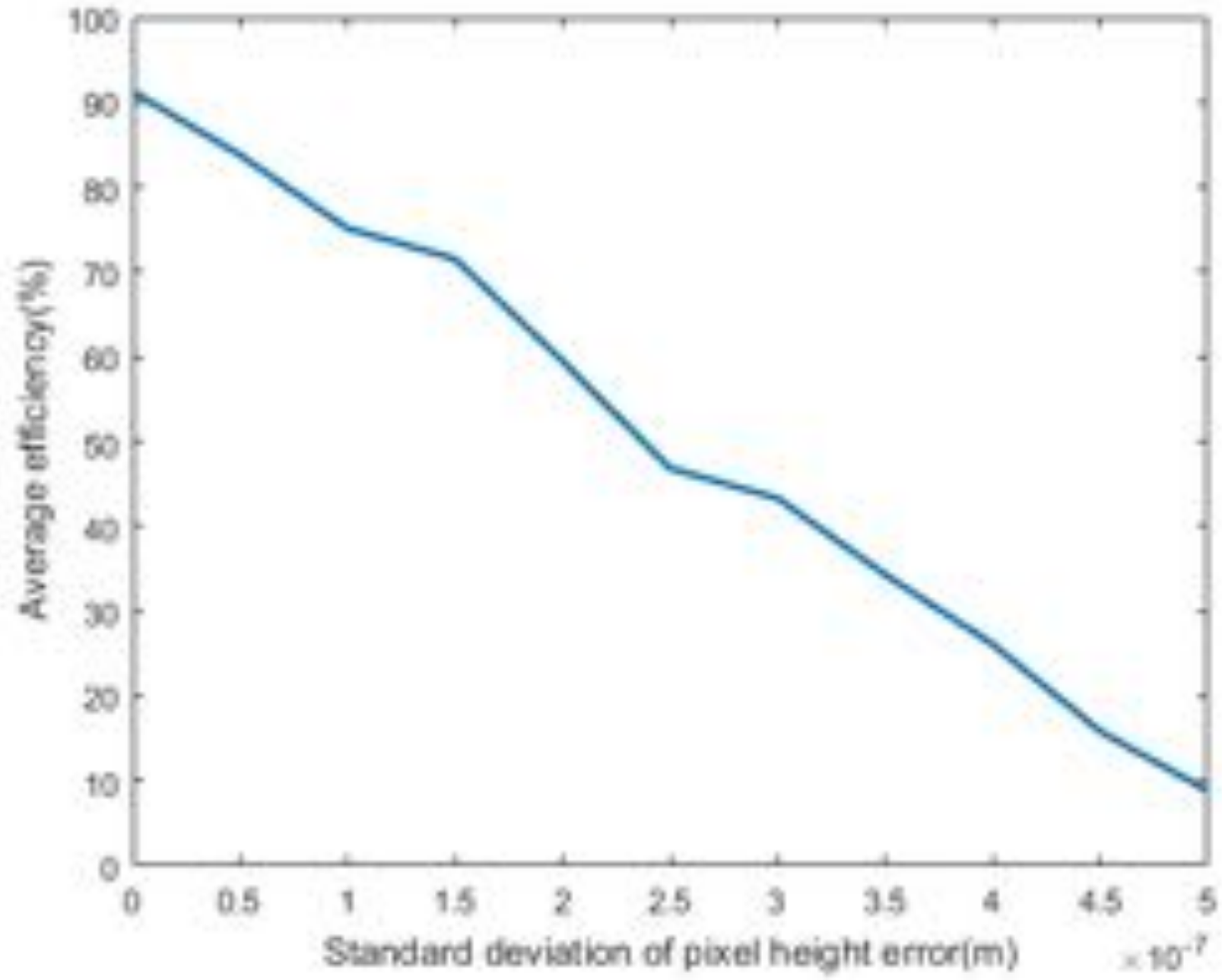

***Fig. S8:*** *Standard deviation based random pixel-height error analysis for the MDL design.*

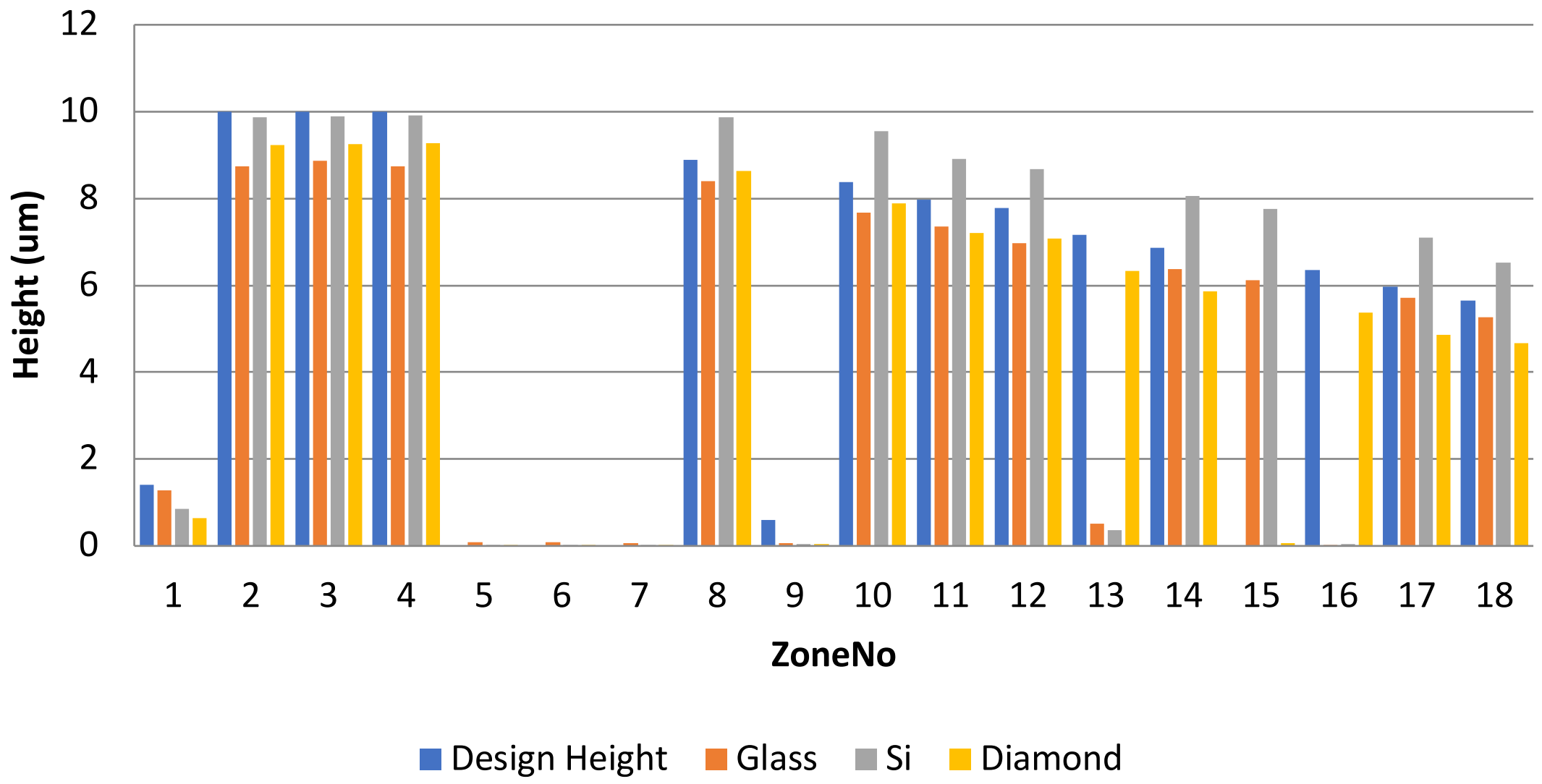

***Fig. S9:*** *Measured and design height of 18 randomly selected zones of the fabricated MDLs*

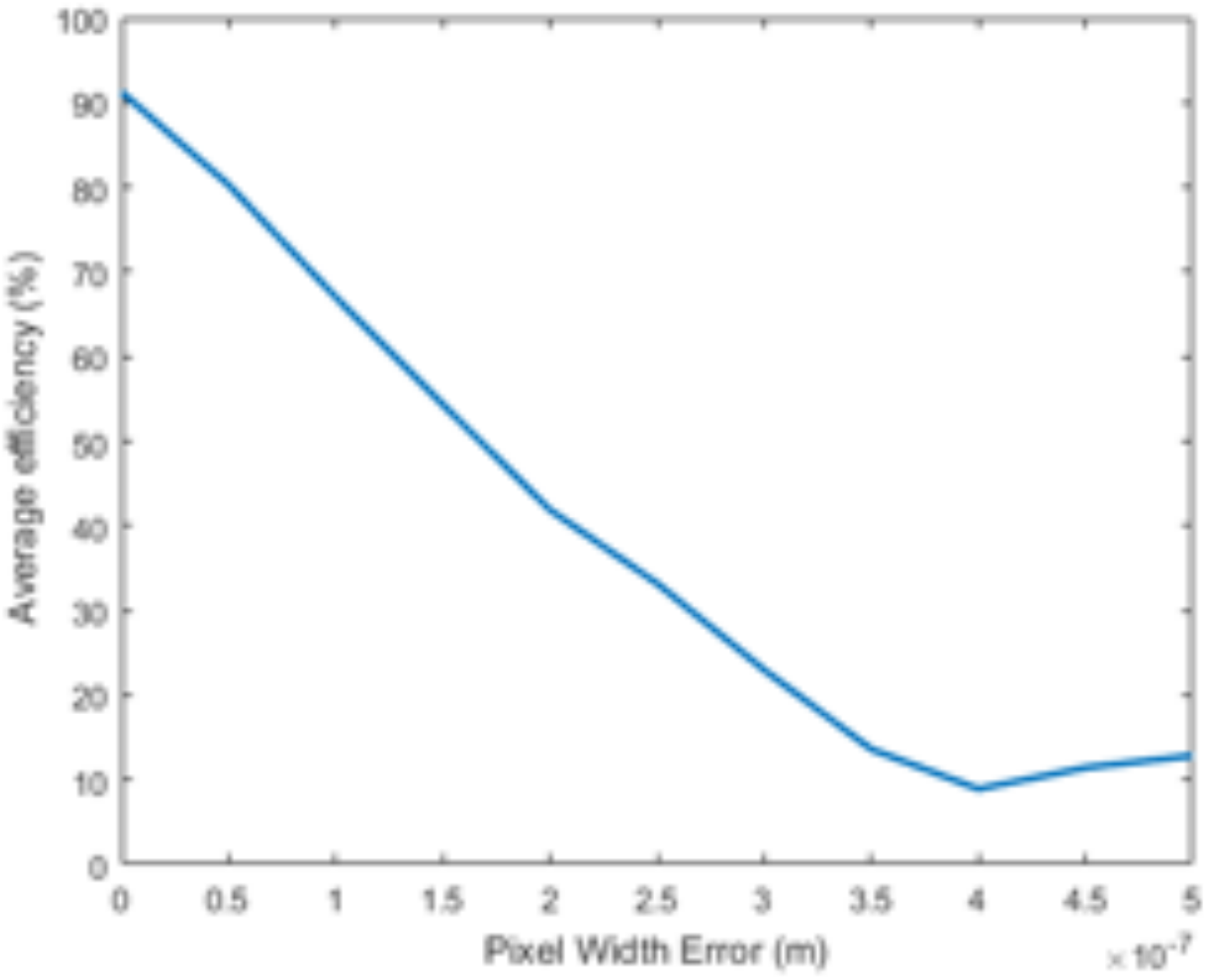

***Fig. S10:*** *Pixel-width error analysis for the MDL design.*

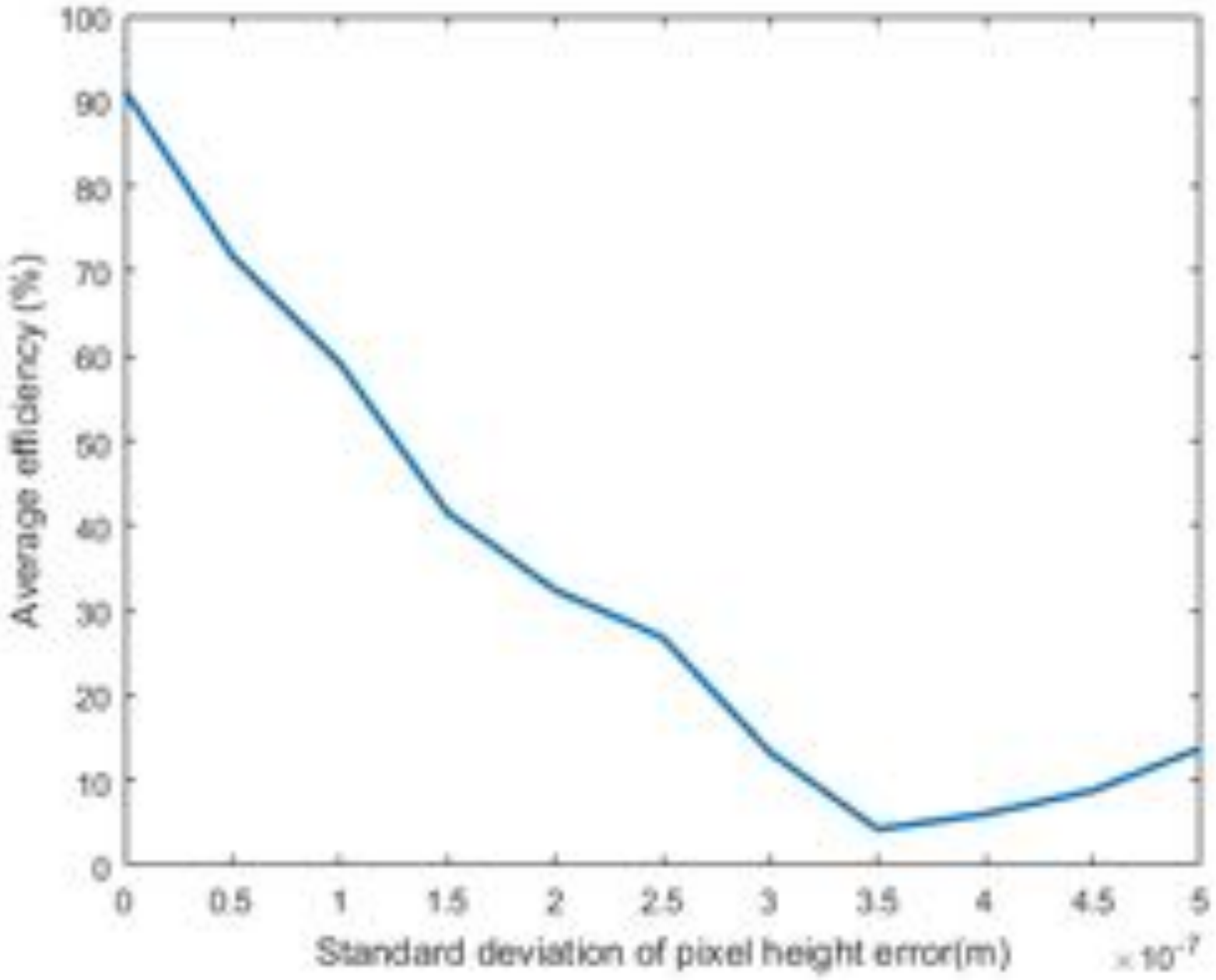

***Fig. S11:*** *Standard deviation based random pixel-height error analysis for the MDL design with a pixel width error = 250 nm.*

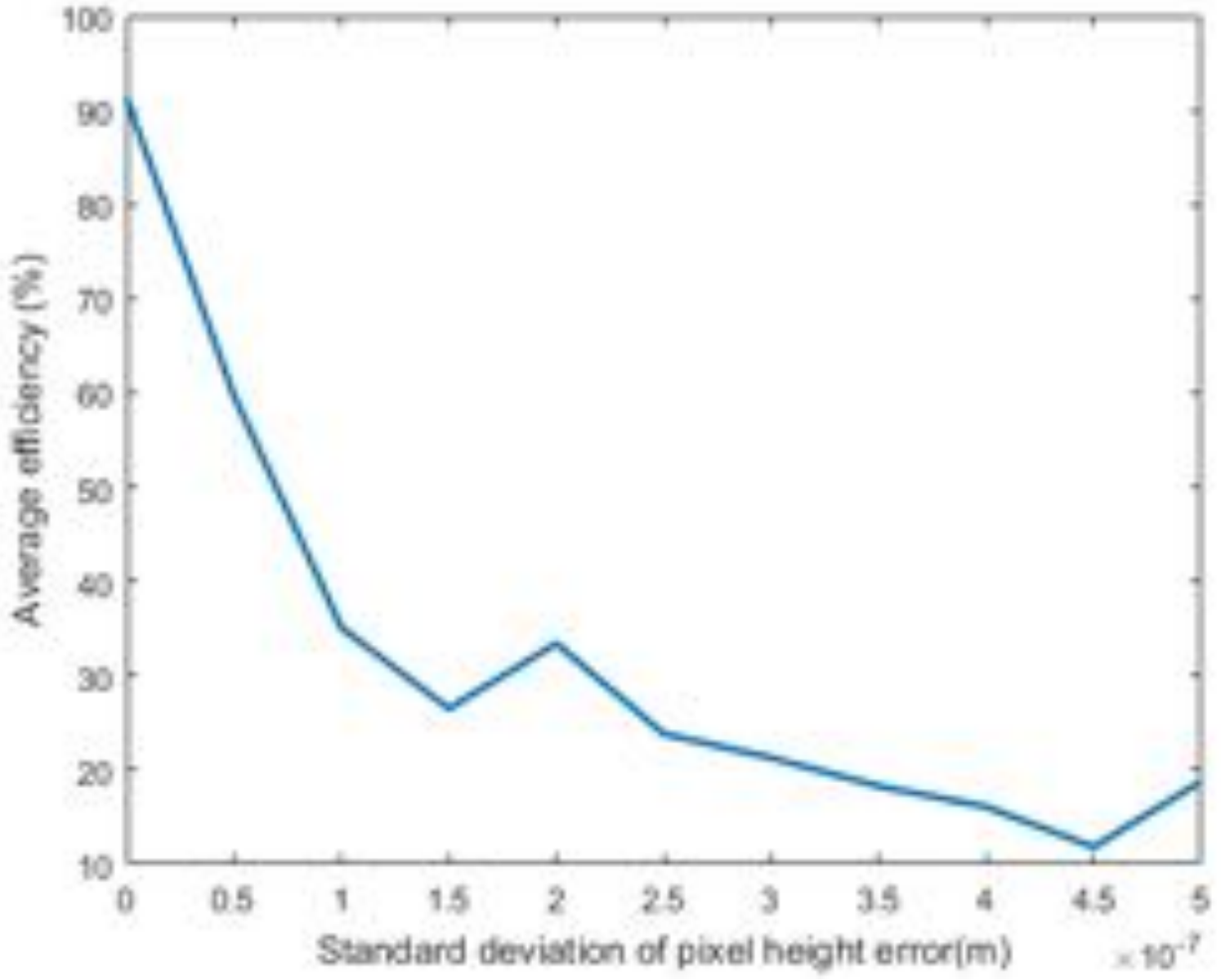

***Fig. S12:*** *Standard deviation based random pixel-height error analysis for the MDL design with a pixel width error = 400 nm*

We assumed a constant refractive index (n) and absorption coefficient (k) of the photoresist (AZ 9260) as 1.61 and 0 respectively for all wavelengths below λ = 1.7 μm respectively while designing the MDL. This is not very accurate from the perspective that optical constants will vary with the wavelength. Hence, we attribute that apart from the pixel height error, an error in the refractive index could explain the discrepancy in the measured PSFs especially below λ = 1.7 μm. To investigate this issue in more detail, we took one exemplary measured PSF at λ = 0.65 μm where significant side lobes were observed during the measurements and tried to simulate our MDL design at the same wavelength but with a slight change in refractive index (from n = 1.57 to n = 1.65). From figure 1(a), we observe that the efficiency value (even at a single wavelength) falls by ~32% within just ±3% of the change in the refractive index alone (with any change in the absorption coefficient). In fact, it is be observed from Fig. 1(b-f), that side lobes start to occur when we slightly deviate from our assumed refractive index value (n = 1.61). The simulation was carried out; based on the assumption that no error in pixel heights was incurred during the fabrication process. Coupled with the fact, that errors propagating from the pixel height and pixel width in the design, it can be assumed that the disagreement in measured PSFs (especially for wavelengths below 1.7 μm) can be attributed to this combined phenomenon.

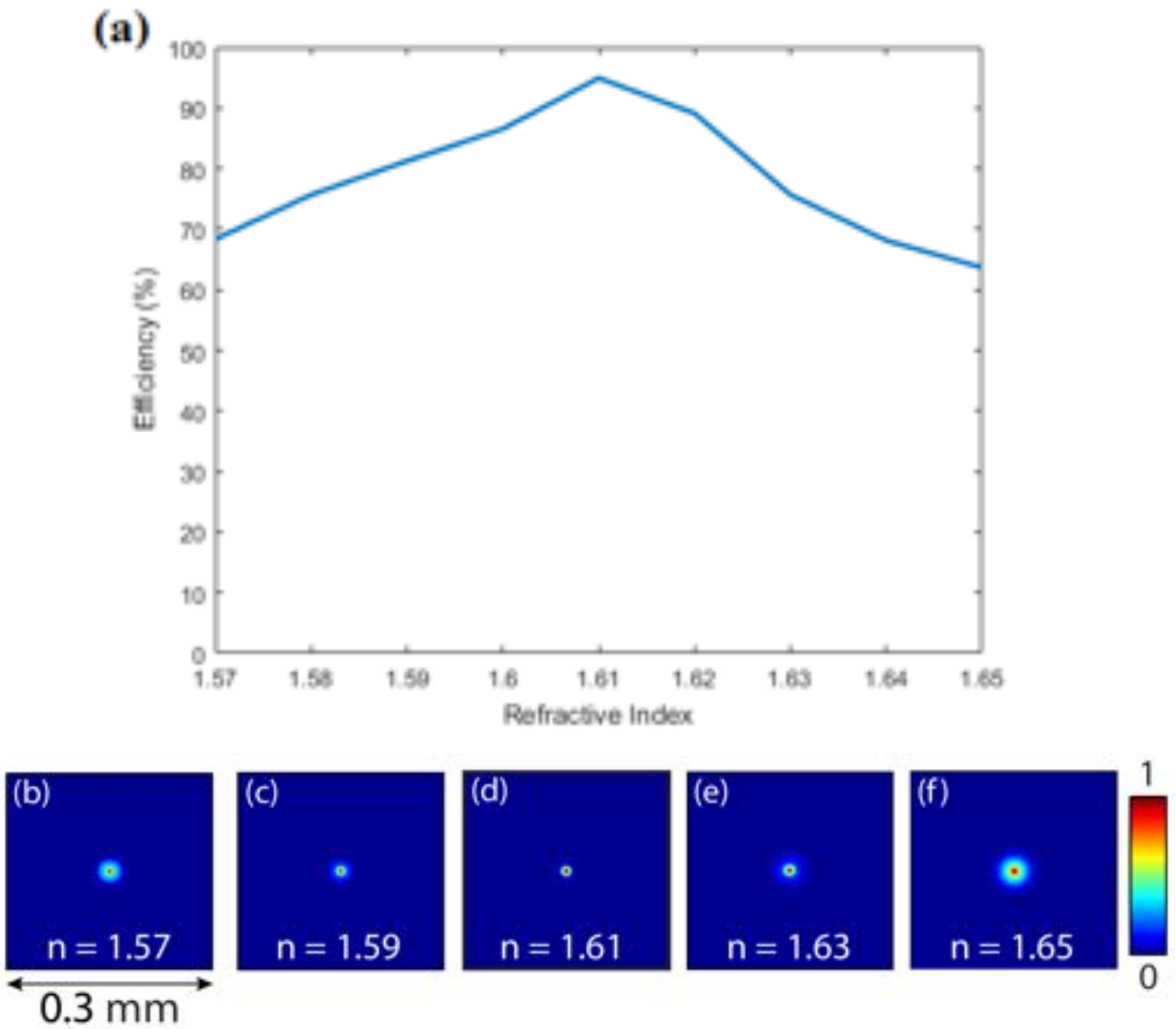

***Fig. S13: (a)** Effect of refractive index on the simulated PSF at* λ = 0.65 μm. *Simulated PSF when **(b)** n = 1.57 **(c)** n = 1.59 **(d)** n = 161 **(e)** n = 1.63 and **(f)** n = 1.65.*

Another corollary observation was that the PSFs also did tend to get bigger for some change in refractive index values which could also explain the discrepancy in the FWHM values from λ = 2 μm to λ = 7 μm.

## 2. Aberrations analysis for measured wavefront of the MDL

### 2.1 Aberrations analysis for measured wavefront of ultra-broadband MDL in the wavelength range 400-950 nm

The following optical setup, shown in Fig. S10 was constructed to measure the aberrations of the ultra-broadband multi-level diffractive lens (UBB MDL) using a Shack Hartmann wavefront sensor. The incident beam from a SuperK EXTREME EXW-6 source [1] was filtered and split into two beams by the SuperK VARIA filter [2] for wavelengths 400-800 nm and a SuperK SELECT filter [3] for wavelengths 800-950 nm. The beams were expanded, collimated, and directed using a series of optical mirrors, iris and flip mirror toward the measurement setup. An iris was placed in the path of the beam to limit the beam diameter. A flip mirror was used to align both the visible and NIR beams to the same path. The final beam was set to have a diameter of 25.4 mm. This beam was incident on the UBB MDL. A Shack-Hartmann wavefront sensor from Thorlabs (WFS 150-7AR) [4] was placed 99.45 mm (UBB MDL to outer rim of the wavefront sensor) behind the test lens [5]. The distance between the outer rim of the WFS 150-7AR to the sensor plane is 13.6 mm, making the total distance between the MDL being tested and the wavefront sensor to be 113.05 mm. The incoming collimated beam converges to a focal spot at ~ 18 mm behind the MDL and then diverges, before contacting the wavefront sensor. Hence, it is expected that the sensor should show a diverging wavefront, which is confirmed by the wavefront measurement. We selected between 3.3 and 4.5 mm as the diameter of the pupil of the WFS 150-7AR based on the illumination wavelength so as to fit the pupil appropriately to the beam width, making sure to not over-sample or under-sample. This was confirmed by using the auto-selection of pupil diameter to beam width feature available in the sensor software. Hence, this ensures that neither is the beam clipped, nor oversampled. Prior to the measurements with the UBB MDL, the radius of curvature of the expanded and collimated and hence plane wave from our sources was confirmed to be in the order of >30 m, signifying a proper plane wave illumination. While measuring the wavefronts the distance between the UBB MDL and the WFS 150-7AR was never changed. Only the SuperK filters were switched in between the visible and NIR measurements.

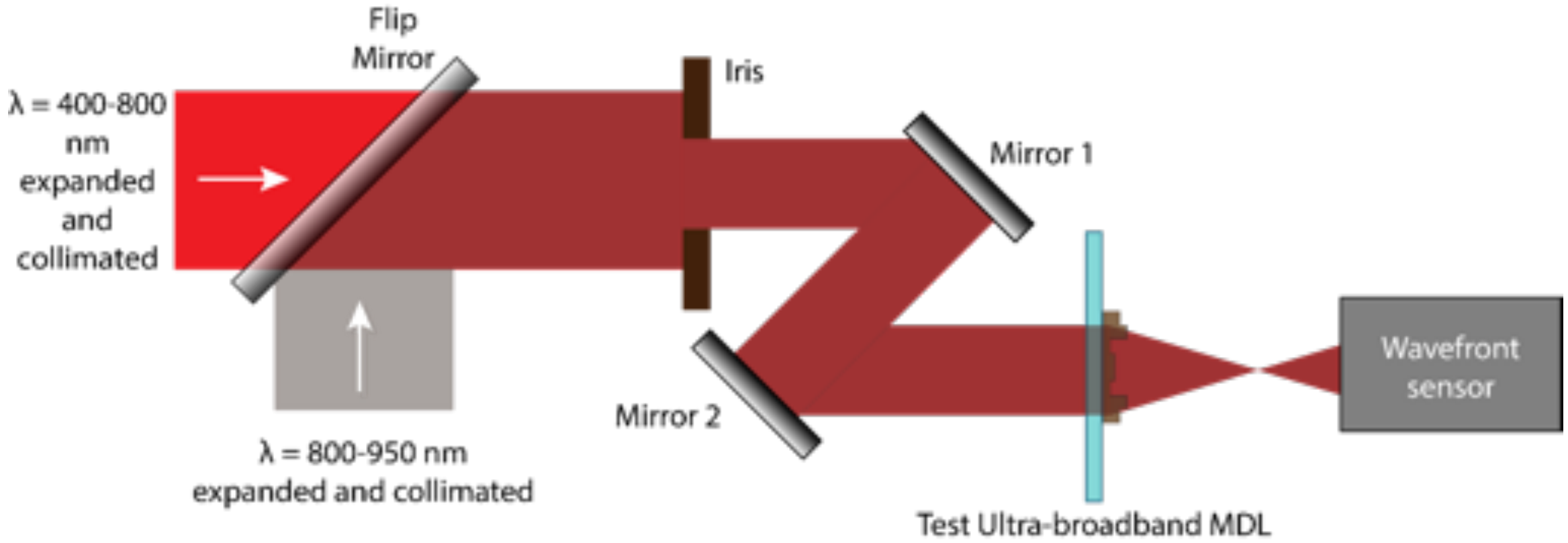

***Fig. S14:*** *Optical setup used to measure aberrations of a test ultra-broadband multi-level diffractive lens. The wavefront sensor is illuminated by a diverging wavefront produced after a collimated beam is brought into focus ~ 18 mm behind the MDL. A series of mirrors and iris are used to relay the beam from the source to the MDL. A flip mirror is used to divert the NIR (800-950 nm) beam and align it in the path of the visible (400-800 nm) beam.*

First, we used a stock refractive singlet lens (plano-convex lens, Thorlabs) to test and calibrate the setup. The alignment is confirmed to be good from the results obtained. Focal length of the lens as obtained from the SH WFS is checked with the know focal length. The known focal length of the lens was 200 mm as specified by the manufacturer. We measured the focal length to be 200.86 mm under the broadband illumination (450-850 nm) and 200.86 mm under 600 nm illumination based on the data gathered from the SH WFS. This step serves as a calibration step and the values of the stock lens serve as ground truth. Next, the MDL was placed in the path of the illuminating beams and the measurements recorded under different illumination conditions. In the case of the narrowband illuminations in the visible range, i.e. for 450, 500, 550, 600, 650, 700, 750 and 800nm, the bandwidth was set to be 50 nm and for the NIR range. *i.e*,. 850, 900 and 950 nm was set to be 15 nm as permitted by the SuperK VARIA and SELECT filters, respectively. The measured focal lengths of the MDL were 0.988 mm under the broadband illumination (450-850 nm) and 1.06 mm under 600 nm illumination based on the data gathered from the SH WFS. The wavefront for the stock lens and the MDL under various illumination conditions are presented in Fig. S11.

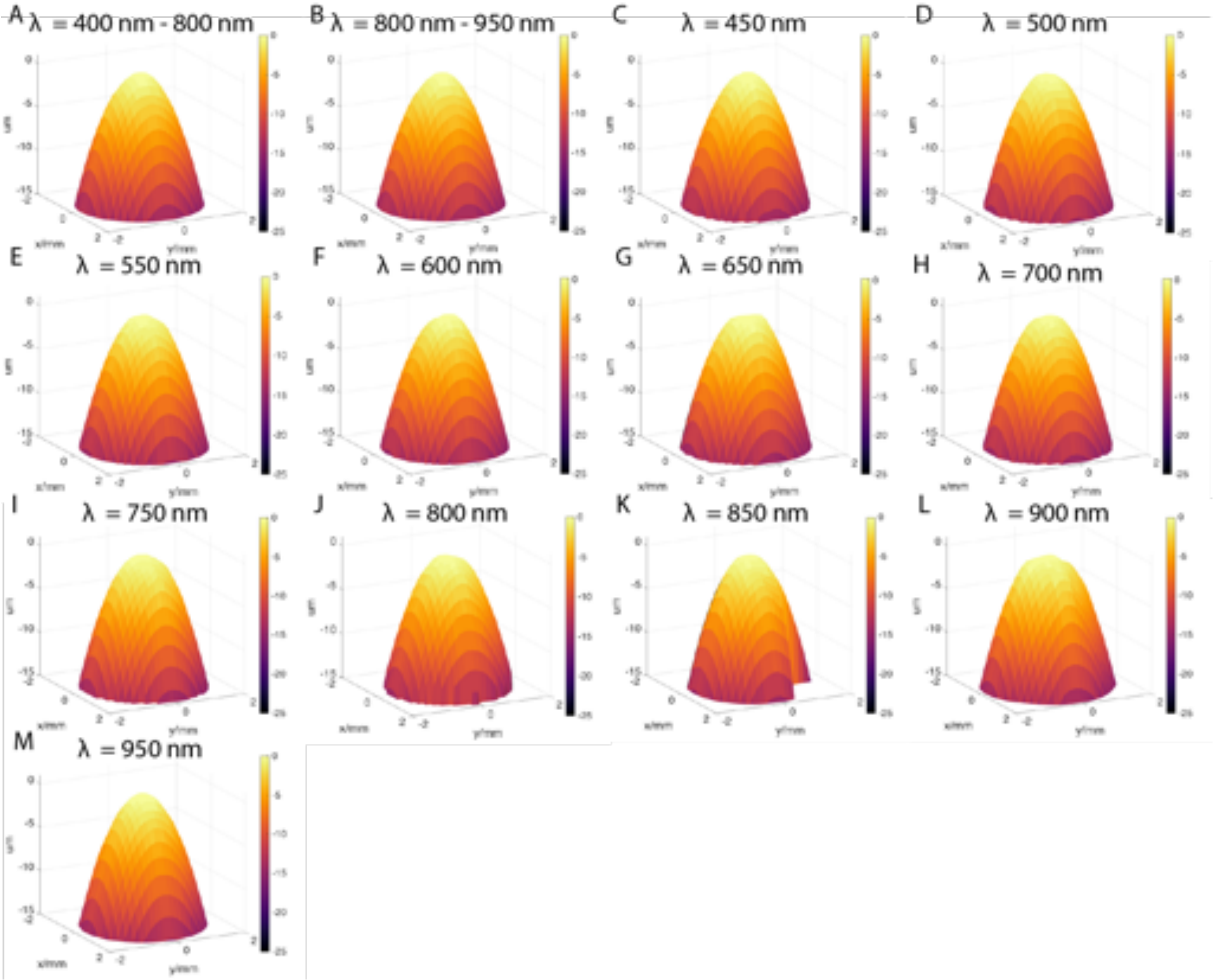

***Fig. S15:****Measured wavefront of the ultra-broadband MDL under (A) broadband visible (400-800 nm) illumination and (B) broadband NIR (800-950 nm) illuminations and (C-M) narrowband (450, 500, 550, 600, 650, 700, 750, 800, 850, 900 and 950 nm respectively) illuminations. The bandwidth in each case for (C-J) was 50 nm and in each case for (K-M) was 15 nm.*

**Table S1.** Measured aberration values for the test ultra-broadband MDL lens under different illumination conditions

| Aberration | | | | Test Ultra-broadband MDL Zernike coefficient value (µm) | | | | | | | | | |
|---|---|---|---|---|---|---|---|---|---|---|---|---|---|
| | λ = 400-800 nm | λ = 800-950 nm | λ = 450 nm | λ = 500 nm | λ = 550 nm | λ = 600 nm | λ = 650 nm | λ = 700 nm | λ = 750 nm | λ = 800 nm | λ = 850 nm | λ = 900 nm | λ = 950 nm |
| Piston | 4.128 | 6.855 | 4.426 | 4.509 | 5.141 | 4.894 | 4.728 | 4.59 | 4.481 | 4.177 | 5.378 | 6.715 | 5.196 |
| Tip Y | 0.006 | 0.063 | -0.076 | 0.046 | 0.007 | 0.053 | 0.039 | -0.039 | 0.017 | 0.013 | -0.067 | 0.173 | -0.164 |
| Tilt X | 0.012 | -0.004 | 0.024 | 0.081 | -0.185 | -0.069 | 0.092 | 0.088 | 0.165 | -0.066 | 0.178 | -0.032 | 0.201 |
| Astig. ±45° | 0.021 | 0.049 | 0.035 | 0.037 | 0.037 | 0.017 | 0.046 | 0.03 | 0.029 | 0.031 | 0.028 | 0.047 | 0.037 |
| Defocus | -4.143 | -7.023 | -4.651 | -4.674 | -5.024 | -4.942 | -4.923 | -4.689 | -4.67 | -4.239 | -5.577 | -6.891 | -5.564 |
| Astig. 0/90° | 0.001 | 0 | 0.004 | 0.012 | 0.001 | 0.006 | -0.002 | 0.002 | 0.011 | 0.001 | 0.036 | 0.012 | -0.006 |
| Trefoil Y | -0.003 | -0.003 | 0.011 | 0.003 | 0.001 | 0.001 | -0.003 | 0.001 | -0.007 | 0.007 | 0.003 | -0.004 | 0.006 |
| Coma X | -0.001 | -0.002 | 0.026 | -0.013 | -0.002 | -0.001 | 0.002 | 0.001 | 0.001 | -0.002 | 0.051 | -0.021 | 0.042 |
| Coma Y | -0.006 | 0.018 | 0.076 | -0.001 | 0.002 | 0.01 | 0.021 | 0.019 | -0.007 | 0.033 | -0.049 | 0.005 | 0.012 |
| Trefoil X | -0.011 | -0.01 | 0.012 | -0.003 | -0.004 | -0.005 | -0.008 | -0.005 | -0.009 | 0.011 | 0.02 | -0.006 | 0.005 |
| Tetrafoil Y | -0.001 | -0.006 | 0.004 | 0 | 0 | -0.001 | -0.006 | 0 | 0.003 | 0.002 | -0.006 | -0.006 | -0.008 |
| Sec. Astig. Y | 0.004 | 0 | 0.014 | -0.002 | 0.001 | 0.006 | -0.002 | -0.004 | 0.005 | -0.006 | -0.027 | 0.003 | 0.008 |
| Spher. Aberr. 3rd O | 0.001 | 0.053 | 0.059 | 0.008 | 0.024 | 0.031 | 0.021 | 0.018 | -0.015 | 0.023 | 0.063 | -0.014 | 0.199 |
| Sec. Astig. X | 0.003 | 0.004 | 0.043 | -0.002 | 0.002 | -0.001 | -0.002 | -0.004 | 0.001 | 0.009 | -0.024 | 0.003 | 0.002 |
| Terafoil X | -0.011 | 0.006 | -0.008 | -0.002 | -0.001 | -0.001 | -0.003 | -0.008 | -0.003 | 0.006 | -0.008 | 0.016 | 0.035 |

## 2.2 Measurement of focal length using Shack Hartmann WFS 150-7AR wavefront sensor for the test ultra-broadband MDL in the wavelength range 400-950 nm

The measurement follows a well-known procedure [6, 7] that calculates the focal length of a test lens using the radius of curvature (ROC) measured from the WFS and the recorded distance between the test lens and the WFS. This is expressed mathematically as follows:

$$\frac{1}{f} = \frac{1}{f+Z-Z_o} + \frac{1}{R+L} \quad (1)$$

where $f$ is the focal length of the lens being tested, $L$ is the distance between the test lens and the wavefront sensor and $R$ is the radius of curvature (ROC) of the wavefront produced by the lens as measured by the wavefront sensor, $Z$ is the distance between the input object and the test lens and $Z_o$ is the distance between the lens and the collimated source. In our case, we only use an incoming collimated beam hence $Z_o \rightarrow \infty$. It is to be also noted that $R$ is the summation of the distance between the test lens and the outer rim of the WFS and the distance between the outer rim to the sensor. The values of these distances have been mentioned above.

### 3. Resolution chart images in visible and NIR

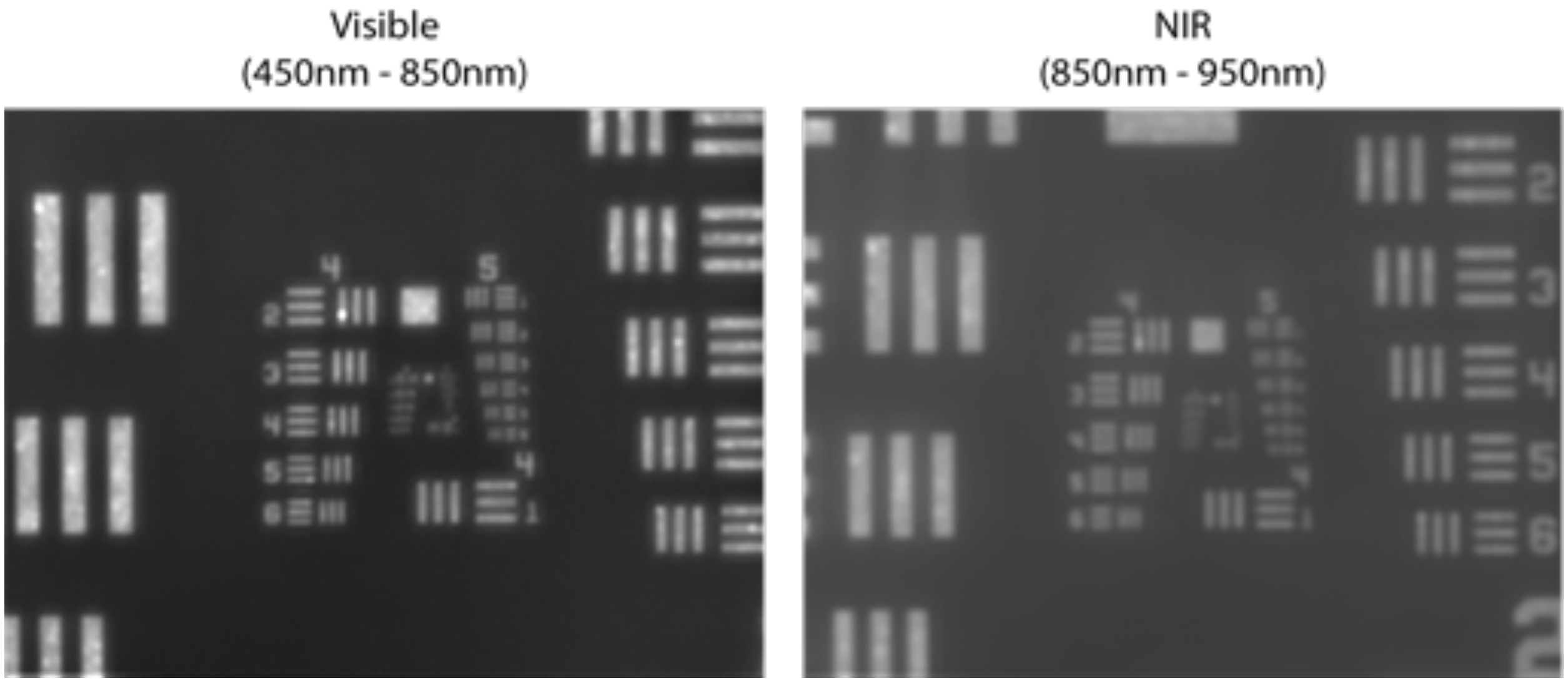

***Fig. S16:*** *Resolution chart imaging in visible (450nm – 850nm) and near NIR (850nm – 950nm) taken with glass-substrate MDL at higher magnification.*

### 4. List of Supplementary Videos

Table S2: List of supplementary video files.

| Filename | MDL substrate | Illumination / Object | Sensor |
|---|---|---|---|
| Supplementary Video 1 | Glass | Sunlight with NIR-cut filter | Si CMOS |
| Supplementary Video 2 | Glass | Sunlight with Vis-cut filter | Si CMOS |
| Supplementary Video 3 | Glass | White LED | Si CMOS |
| Supplementary Video 4 | Glass | NIR (850nm) LED flashlight | Si CMOS |
| Supplementary Video 5 | Si | Heated resistor coil | FLIR Tau2 |
| Supplementary Video 6 | CVD Diamond | Sunlight with NIR-cut filter | Si CMOS |
| Supplementary Video 7 | CVD Diamond | Sunlight with Vis-cut filter | Si CMOS |
| Supplementary Video 8 | CVD Diamond | Various thermal objects | FLIR Tau2 |

The Si CMOS refers to the monochrome sensor DMM 27UP031-ML from the Imaging Source. Supplementary videos 1, 2, and 8 were created by merging together two (1,2) or three (8) separate files using the free web-service provided by clideo.com.